\documentclass[onecolumn]{aastex631}

\usepackage{subfigure}

\newcommand{\code}[1]{{\color{blue} \texttt{#1}}}
\newcommand{\pack}[1]{{\color{magenta} \texttt{#1}}}
\newcommand{\oiii}{\lbrack OIII\rbrack}

\begin{document}

\title{Direct imaging of AGN outflows and their origin with the 23 m Large Binocular Telescope}

\author{J. W. Isbell}
\correspondingauthor{J. W. Isbell}
\email{jwisbell@arizona.edu}
\affiliation{Department of Astronomy and Steward Observatory, The University of Arizona, 933 North Cherry Ave, Tucson, AZ, 85721, USA}
\affiliation{Max-Planck-Institut für Astronomie, Königstuhl 17, Heidelberg 69117, Germany}

\author{S. Ertel}
\affiliation{Department of Astronomy and Steward Observatory, The University of Arizona, 933 North Cherry Ave, Tucson, AZ, 85721, USA}
\affiliation{Large Binocular Telescope Observatory, The University of Arizona, 933 North Cherry Ave, Tucson, AZ, 85721, USA}

\author{J.-U. Pott}
\affiliation{Max-Planck-Institut für Astronomie, Königstuhl 17, Heidelberg 69117, Germany}

\author{G. Weigelt}
\affiliation{Max-Planck-Institut f\"ur Radioastronomie, Auf dem H\"ugel 69, D-53121 Bonn, Germany}

\author{M. Stalevski}
\affiliation{Astronomical Observatory, Volgina 7, 11060 Belgrade, Serbia}
\affiliation{Sterrenkundig Observatorium, Universiteit Gent, Krijgslaan 281-S9, Gent, 9000, Belgium}

\author{J. Leftley}
\affiliation{Laboratoire Lagrange, Universit\'e C\^ote d'Azur, Observatoire de la C\^ote d'Azur, CNRS, Boulevard de l'Observatoire, CS 34229, 06304 Nice Cedex 4, France}

\author{W. Jaffe}
\affiliation{Leiden Observatory, Leiden University, Niels Bohrweg 2, NL-2333 CA Leiden, The Netherlands}

\author{R. G. Petrov}
\affiliation{Laboratoire Lagrange, Universit\'e C\^ote d'Azur, Observatoire de la C\^ote d'Azur, CNRS, Boulevard de l'Observatoire, CS 34229, 06304 Nice Cedex 4, France}

\author{N. Moszczynski}
\affiliation{LESIA, Observatoire de Paris, Université PSL, CNRS, Sorbonne Université, Université de Paris Cité, 5 place Jules Janssen, 92190
Meudon, France}

\author{P. Vermot}
\affiliation{LESIA, Observatoire de Paris, Université PSL, CNRS, Sorbonne Université, Université de Paris Cité, 5 place Jules Janssen, 92190
Meudon, France}

\author{P. Hinz}
\affiliation{University of California, Santa Cruz, 1156 High St, Santa Cruz, CA 95064, USA}

\author{L. Burtscher}
\affiliation{Umweltinstitut M\"unchen, Goethestraße 20, 80336 München, Germany}

\author{V. G\'amez Rosas}
\affiliation{Leiden Observatory, Leiden University, Niels Bohrweg 2, NL-2333 CA Leiden, The Netherlands}

\author{A. Becker}
\affiliation{Large Binocular Telescope Observatory, The University of Arizona, 933 North Cherry Ave, Tucson, AZ, 85721, USA}

\author{J. Carlson}
\affiliation{Large Binocular Telescope Observatory, The University of Arizona, 933 North Cherry Ave, Tucson, AZ, 85721, USA}

\author{V. Faramaz-Gorka}
\affiliation{Department of Astronomy and Steward Observatory, The University of Arizona, 933 North Cherry Ave, Tucson, AZ, 85721, USA}

\author{W. F. Hoffmann}
\affiliation{Department of Astronomy and Steward Observatory, The University of Arizona, 933 North Cherry Ave, Tucson, AZ, 85721, USA}

\author{J. Leisenring}
\affiliation{Department of Astronomy and Steward Observatory, The University of Arizona, 933 North Cherry Ave, Tucson, AZ, 85721, USA}

\author{J. Power}
\affiliation{Large Binocular Telescope Observatory, The University of Arizona, 933 North Cherry Ave, Tucson, AZ, 85721, USA}

\author{K. Wagner}
\affiliation{Department of Astronomy and Steward Observatory, The University of Arizona, 933 North Cherry Ave, Tucson, AZ, 85721, USA}



\begin{abstract}
Active galactic nuclei (AGNs) are a key component of galaxy evolution due to feedback on the host from its supermassive black hole. The morphology of warm, in- and outflowing dusty material can reveal the nature of the onset of feedback, AGN feeding, and the unified model of AGN. Here we use the Large Binocular Telescope Interferometer (LBTI) to image the dense, obscuring disk and extended dusty outflow region of NGC~1068.  In Fizeau imaging mode the LBTI synthesizes the equivalent resolution of a 22.8 m telescope.  The 8.7~$\mu$m Fizeau images of NGC~1068 {have an effective resolution of $47\times90$ mas ($3.3\times6.2$ pc)} in a 5" field of view after performing PSF deconvolution techniques described here.
This is the only extragalactic source to be Fizeau imaged using the LBTI, and the images bridge the scales measured with the Very Large Telescope Interferometer (VLTI; 0.5-5~pc) and those of single telescopes such as JWST and Keck ($>15$~pc). The images detect and spatially resolve the low surface brightness mid-infrared (MIR) features in the AGN disk/wind region that are over-resolved by the VLTI.  The images show strong correlation between MIR dust emission and near-infrared (NIR) emission of highly excited atomic lines observed by SINFONI.
Such LBTI imaging is a precursor to infrared imaging using the upcoming generation of extremely large telescopes, with angular resolutions up to 6x better than JWST, the largest space telescope in orbit.
\end{abstract}



\section*{} \label{sec:intro}
\part*{Main}
The study of AGN at high spatial resolution aims to test the Unified Model of AGN\citep{antonucci1993}, understand the AGN lifecycle, and resolve the onset of AGN feedback -- crucial information for studies of AGN at high redshift and within cosmological simulations. Dusty outflows are of particular interest because they dominate the energy budget of AGN feedback, especially in highly accreting AGN (e.g., quasars; \citet{fabian2012}). The winds are thought to be driven by radiation pressure from high-energy photons in the vicinity of the SMBH \citep[e.g.,][]{wada2012,wada2016, williamson2019,leftley2019, williamson2020}. In simulations, anisotropic radiation from the accretion disk \citep{netzer1987} drives the dust primarily in the polar direction, but the outflows are further impacted by and impact inflowing material in a so-called fountain flow \citep{wada2012}. Polar extended dust has been commonly observed on parsec scales in a number of AGN with VLTI/MIDI \citep{burtscher2013,lopezgonzaga2016} and studied more extensively in two nearby AGN with VLTI/MATISSE \citep{isbell2022,isbell2023,gamezrosas2022}. However, how the AGN and its winds are fueled and how the polar ejected material interacts with the host galaxy remain open questions.

NGC 1068 is an important laboratory for the understanding of AGN feeding and feedback. Located only 14.4 Mpc away, it is the prototypical Seyfert 2 AGN \citep{antonucci1993}; it led to the development of the Unified AGN model. Its proximity and brightness have allowed for extremely high spatial resolution imaging of its nucleus in the near-infrared \citep[NIR;][]{gravitycollaboration2020}, MIR \citep{gamezrosas2022}, sub-mm \citep{garciaburillo2016,impellizzeri2019,garcia-burillo2019}, and radio \citep{gallimore2004, mutie2024}. The sub-mm  observations probe cold dust and dense molecular gas on similar scales as the LBTI. 
{Interpretation of the connections between dusty, molecular material, hot ionized outflows, and the AGN itself is complicated by interactions with the ``bent'' radio jet\citep{may2017}}, which leaves the core oriented approximately North-South, but is seen at large scales to be oriented northeast-southwest \citep[e.g., ][]{mutie2024}. Additionally, high-resolution NIR/MIR imaging at the VLTI with the 8.1 m Unit Telescopes resolves out structures larger than $\approx 4$~pc at 8.7~$\mu$m, so the link between the small-scale and galactic structures is tenuous. Measurements on scales intermediate to the existing 8.1 m single-dish observations and the 34 m VLTI {shortest unit telescope (UT)} baseline are necessary to understand how material is delivered to the accretion disk and to the nascent outflows.

The LBTI \citep{hinz2016, ertel2020} co-phases and interferometrically combines the beams from the two 8.4~m mirrors of the Large Binocular Telescope (LBT), separated 14.4~m.  
In Fizeau imaging, this results in a virtual telescope aperture with a maximum baseline of 22.8~m. 
While the Auxiliary Telescopes (ATs) at the VLTI offer similar baselines, their 1.8 m diameters provide limited sensitivity compared to the LBTI. 
The LBTI thus provides complementary data to the existing VLTI observations of NGC 1068. 
Open phase loop Fizeau interferometric imaging is the simplest interferometric mode of the LBTI. 
It has great potential for the high-fidelity, wide-field ($\sim$5\,arcsec), intermediate-contrast imaging of faint near-/mid-IR targets or targets that are too extended for phase tracking (examples include Solar System bodies, AGNs, ultra-luminous infrared galaxies, and massive and evolved stars). 
The mode, however, has not been actively developed since its successful observations of Io \citep{leisenring2014,conrad2015,dekleer2017}.
Open loop Fizeau imaging at the LBTI is poised to bridge the gap between high-dynamic-range single-dish observations with e.g., JWST, and low-dynamic-range high-resolution interferometric images with e.g., VLTI and CHARA. Its 22.8 m resolution is comparable to the upcoming generation of $30$~m class telescopes, and can serve as a key testing ground for science cases and techniques. 

{ In this paper, we use open loop Fizeau imaging with the LBTI on an extragalactic target, providing spatial constraints on the structure and heating of the nuclear outflow and dust torus in NGC 1068. These observations reveal low surface brightness features at unprobed scales intermediate to single-dish telescopes and long baseline infrared interferometry.}

\part*{Observations and Data Reduction} \label{sec:obs}
On UTC 6 November 2022 we observed NGC 1068, 1 Tau, and Procyon using the NOMIC instrument \citep{hoffmann2014} on the LBTI in the Open-Loop Fizeau mode. We used the W08699-9\_122	filter ($\lambda_c = 8.7~\mu$m, range $[8.13, 9.35]~\mu$m) and took 20 nodding cycles on the target and calibrators, obtaining $59^{\circ}$~field rotation. Before and after the science target we observed 1 Tau and Procyon to serve as flux and point spread function (PSF) calibrators. We later selected images via ``lucky fringing,'' resulting in an exposure time of $200 \times 0.0137$~s$= 2.74$~s per cycle or $54.8$~s total exposure on NGC 1068. The resulting corotated and stacked target images and PSF calibrator images are each shown in Fig. 1. See Methods \S1 for details on frame selection, PSF calibration, and flux calibration.

We deconvolved the calibrator PSF from the science target PSF in order to recover the underlying NGC 1068 source flux distribution. While this had previously been done for Io \citep{leisenring2014} with the LMIRCam detector, NGC 1068 is much fainter and extended, and several changes were made to the AO system of the LBT \citep{pinna2016} in the years following the publication of those data. We therefore performed image deconvolution in three different ways (see Methods \S2), finding good agreement between them. We show the results of deconvolution using the Richardson-Lucy (R-L) method as in \citep{leisenring2014} in Fig. 2. Other deconvolutions are shown in Extended Data (ED). We estimate the resulting resolution using the southernmost and northernmost point sources in the image, measuring FWHM $47\times 90$~mas after deconvolution. In this article, we focus on the R-L deconvolved images of NGC 1068 because of the robust legacy and fidelity of the method.

\part*{The Dusty Wind}
The circumnuclear region in NGC 1068 has been extensively studied in high resolution at numerous wavelengths. 
We compare the recovered 8.7~$\mu$m morphology at 47 mas (3.3 pc) resolution to existing radio maps (at 5 GHz with $105\times 51$~mas resolution; \citet{gallimore2004}), optical images (\lbrack OIII5007\rbrack~emission~with 52 mas resolution; \citet{macchetto1994}), and NIR/MIR observations using single-dish deconvolution ($\sim 100$~mas resolution; \citet{bock2000}) and interferometry \citep[][74 mas and 3 mas resolution, respectively]{weigelt2004, gamezrosas2022}. 
We show this comparison in Fig. 3, specifically showing the 5 GHz radio emission, previous single-dish MIR imaging, \lbrack OIII\rbrack~emission, and VLTI/MATISSE MIR reconstructed image with respect to our recovered MIR structures. 

In the deconvolved images (Fig. 2 and ED Figs. 1, 2, \& 3), we see significant extended flux along the north-south direction as well as a bright point sources to the northeast at 600 mas (45 pc) and to the south at 300 mas (20 pc, called the ``SN'' for southern $N$-band source). The $47\times90$~mas recovered resolution allows us to distinguish substructures near the AGN core. Here we see extensions of flux from the flux peak toward the north and northeast out to $\gtrsim 10$~pc (150 mas). We refer to this northeast extension hereafter as ``NEX'' and the northern extension as ``NEP.'' Between the flux peak and the SN there is a drop in emission we refer to as the ``dark band.'' 

We find morphological agreement with both the radio emission and previous optical-infrared imaging. {The authors in \citet{gallimore2004} locate the SMBH in a structure called S1; similar to positions inferred from maser emission\citep{gallimore2023}, model fitting to MATISSE data\citep{leftley2023}, and morphological correspondence\citep{bock2000,gamezrosas2022} we place the peak of the MIR emission 50 mas north of S1.} We then find that the southernmost and northern 8.7~$\mu$m structures correspond roughly to peaks in the radio emission (called `S2' and `C' in \citet{gallimore2004}, respectively). However, we note in both cases there is an offset of $\approx 50$~mas, with the MIR emission found exterior to the radio.
The northernmost emission is near the `NE' source of \citep{gallimore2004}, but like in \citep{bock2000} we find a $\sim 100$~mas separation between the MIR and 5~GHz emission. 
The 7.9~\micron~emission of \citep{bock2000} includes a peak far to the NE that we do not recover. Assuming the flux of the northeastern component has remained the same since the \citep{bock2000} observations, we would expect a flux density of $\approx 0.68$~mJy/px, which is above the 2$\sigma$ flux level of the deconvolved image. Over the same period, the 5 GHz flux of cloud `C' has been shown to decrease by $\approx 50\%$ \citep{mutie2024}; a similar flux change would put this source below our current detection limit. Indeed, when comparing our 144 mas aperture flux of cloud `C' to the 200~mas aperture flux of \citep{bock2000}, we measure approximately one sixth of the previous value ($100 \pm 6$~mJy at 8.7~\micron~versus $661$~mJy at 7.9~\micron).

We see that the northern structures have counterparts in the \oiii~and NIR emission. Notably, our measured NEX feature traces the opening angle of the ionization cone \citep{macchetto1994}. {SINFONI observations of several atomic emission lines (\lbrack Fe II\rbrack, \lbrack Si IV\rbrack, Pa $\alpha$, and HeI) and molecular hydrogen show a similar morphology to the arc of flux starting from `C' and continuing from the tongue counterclockwise toward `NE'\citep{may2017}. The infrared emission is thus coincident with highly ionized, relatively hot gas. The atomic lines are associated with a shock ``bubble'' being excavated by the jet\citep{may2017, vermot2023}.}

Infrared interferometry also shows morphological similarity to the features near the center of the Fizeau image. In the $K^{\prime}$-band bispectrum speckle interferometry images of \citep{weigelt2004}, significant flux extends from the AGN toward the north $\approx400$~mas. The $K^{\prime}$-band emission corresponds in scale and shape to the northern extended $N$-band flux we image. The authors give a rough flux range of 30–230 mJy; in our northern aperture; we measure $\sim 100$~mJy. The $K^{\prime}$-band flux range is too large to reasonably estimate a dust temperature at this location. 
The VLTI/MATISSE observations\citep{gamezrosas2022} and modeling\citep{leftley2023} indicate the continuation of NEX down to sub-parsec scales. In the modeling work\citep{leftley2023}, this is the southern edge of a dusty (outflow) cone. In the model-independent imaging\citep{gamezrosas2022}, the $N$-band emission also shows an $\times$-shape similar to the ALMA emission \citep{impellizzeri2019} and the edges of the \oiii~emission. {The match in PA of the LBTI Fizeau NEX, the VLTI/MATISSE ``$\times$'', and the ALMA 256 GHz map\citep{impellizzeri2019} strongly indicates that this is the same structure measured at different scales.}

Using apertures with a diameter of 10.1 pc, we measure the flux at two image locations: the flux of the southern component (SN) is $F_{\rm South} = 39.6 \pm 2.4$~mJy, with the aperture centered 25.1 pc to the south of the center; at the same distance to the north we measure $F_{\rm North} = 100.6 \pm 6.2$~mJy. 
Using Eq. 1 from \citep{barvainis1987} and the relation $F(r) =\sigma A T(r)^4$, we calculate an expected dust flux in an aperture with area $A$ as a function of radius (see Methods \S4). The observed values are far above the simple prediction, as is readily seen in Fig. 4. At distances of 20 and 45 pc the flux excesses are associated with 5 GHz features `C' and `NE.' Spectral index measurements of cloud C show steep spectra indicative of optically thin, non-thermal emission between 5 and 21 GHz \citep{mutie2024}. We use the spectral index and 5~GHz integrated flux of `C' to estimate a jet flux at 8.7~$\mu$m of 0.06~mJy. The non-thermal contribution is thus quite small. It is therefore likely that a mixture of non-thermal emission and shock heating is necessary to produce the MIR fluxes. The exact contributions from non-thermal radiation and shock-heated dust are beyond the scope of this work. The correspondence between the different observations indicates that the MIR emission simultaneously traces AGN heated dust, jet-cloud interactions in `C' (and possibly `S2'), and highly ionized jet shocks in the `tongue.' Our observations lack precise astrometry, but independent morphological similarity to the radio, MIR, and \lbrack OIII\rbrack~observations constrains our positions to within a few pixels ($\leq 3$~px$=54$~mas, or one PSF width).

\part*{Constraints on the dusty torus}
The disk+wind model consists of a geometrically thin, optically thick equatorial disk and a conical structure perpendicular to it (usually with hyperbolic or parabolic walls) representing a radiation-driven wind. 
This morphology was recently shown to be a good representation of the VLTI/GRAVITY and VLTI/MATISSE data for NGC 1068 \citep{leftley2023}, but the interferometric data resolved out structures larger than $\sim 5$~pc.
We computed a disk+wind radiative transfer (RT) model out to 15 pc using the \textsc{SKIRT} code \citep{camps2020}. The parameters of the RT modeling are given in Methods \S5. The model was based on the one presented in \citep{isbell2022} with clumpy graphite dust in a hyperboloid cone and graphite-silicate dust in the disk. {Surface brightness profiles of the resulting simulated and observed flux distributions are shown in Fig. 4}. 

Many observed features on $<15$~pc scales (e.g., NEX, the dark band, and SN) and the N-S asymmetry of the large-scale emission are reproduced by the disk+wind model (shown in ED Fig. 4): near the central flux peak we see an edge-brightened outflow cone like NEX; south of the flux peak is a dark absorption band due to the densest part of the disk; and the $\sim 75^{\circ}$ inclined disk out to $\geq 25$~pc causes significant extinction toward the south. The observed morphology, however, is not purely a disk+wind. There is much more flux directly north of the AGN in the images than in the modeling -- this follows the path of the radio jet and is likely emission from jet-cloud interactions. Based on deviations from a centrally heated flux profile, correspondence with radio jet and shock morphology, and 5-21 GHz spectral indices of the sources, we claim that a large fraction of the MIR emission we recover is actually related to the radio jet (either directly or through e.g., shock heating) rather than the radiation-pressure-driven wind.
Through comparisons to RT models as well as to multi-wavelength observations, we are able to spatially distinguish two separate AGN-host interaction channels at 3.4 pc resolution, showing a disk+wind(+jet) morphology. Previous VLT imaging\citep{isbell2021} supported the disk+wind model but did not have sufficient resolution to distinguish jet-related features.

The dusty wind, ionization cone, and radio jet are all presumed to be symmetric above and below the accretion disk. Based on correspondence with the radio data, we place the SMBH at the location marked with a green `$\times$' in Fig. 5, near the bottom of the brightest flux region. If we then assume that similar structure should extend to the north and south, it is immediately apparent that some absorbing medium is present -- likely the disk-like component of the dusty torus. We estimate the optical depth of this dusty structure by relating the flux at an equal distance (accounting for inclination) above and below the SMBH (see Methods \S6). Using the same apertures as above, we find that the southern emitting region is being obscured by an object with an optical depth of $\tau_{\rm 8.7} = 0.41 \pm 0.04$. This obscuring structure is assumed to be the disk-like component of the disk+wind model \citep[e.g.,][]{stalevski2019, leftley2023}. The optical depth of such a structure is expected to increase approaching the SMBH. We have measured the average value at 25.1 pc within a 10.1 pc diameter aperture. Assuming a linear change in $\tau$, we obtain an optical depth gradient of $-0.05~\rm{pc}^{-1} \leq \Delta\tau_{8.7\mu\rm{m}} \leq -0.02~\rm{pc}^{-1}$ when comparing to VLTI/MATISSE $<1$~pc scale measurements\citep{gamezrosas2022}. 

{The dark band between the flux peak in our images and the southernmost emission is inferred to be due to an obscuring dust disk centered on `S1.' VLTI/MATISSE observations revealed a similarly oriented obscuring dust disk\citep[][plotted as yellow ellipses in Fig. 3, panel d; reproduced from \citep{gamezrosas2022}]{gamezrosas2022,leftley2023}. A molecular torus observed with CO(2-1), CO(3-2), and HCO$^{+}$(4-3)\citep{garcia-burillo2019} is oriented approximately perpendicular to the radio jet, similar to the orientation of the obscuring ``dark band'' we and VLTI/MATISSE observe. We estimate that the $75^{\circ}$ inclined obscuring disk has a diameter $\geq 70$~pc in order to obscure SN. This is more than twice as large as the 30~pc CO disk measured by ALMA \citep{garcia-burillo2019}; the discrepancy may indicate that CO traces the densest, inner parts of the disk, but a more extended region still contributes to obscuration.

Finally, using the assumed geometry and inclination of the circumnuclear structures, we estimate a lower limit on the dust density within the obscuring ``dark band'' to be $\rho_{\rm dust} \geq 9\times 10^{-27}~\rm{g}~\rm{cm}^{-3}$. Compared to average ISM gas density, this corresponds to 60\% of the standard ISM dust-gas-ratio of $\approx 1\%$\citep{bohlin1978}. 
A study of 36 X-ray observed AGN yielded dust-to-gas ratios in the range [1\%-100\%] the standard ISM value\citep{esparza2021}. Our measurement falls within this range, but we note that the local density can be much higher if the dust is located within a thin disk as models predict. Future RT modeling is necessary to test the implications of the optical depth gradient on the disk's density profile.

The methods used herein can be applied to other AGN in order to understand the applicability of the disk+wind(+jet) as a unified model of AGN. Based on the measured fringe SNR of these observations, $324.26 \pm 76.17$ for the 20 cycles of NGC 1068 ($10.2\pm3.6$~Jy) and $307.22 \pm 28.17$ for the 7 good cycles of 1 Tau (11.2 Jy), we estimate that we could achieve SNR$\geq 10$ for sources $\geq 0.3$~Jy with roughly one-minute total integration. This is certainly optimistic due to decreased AO performance on faint sources, but repeated observations, more cycles, or fringe tracking on a nearby star could overcome this. We cross-matched the AGN catalog of \citep{veroncetty2010} with the ALLWISE catalog \citep{allwise} to search for sources. Given the 0.3~Jy $N$-band flux cutoff we use $W3$ as a proxy to identify 24 sources which are bright enough in $N$ and have $V < 15$ for AO tracking. This sample of northern AGN includes roughly equal numbers of Seyfert 1 and Seyfert 2 systems, and would allow for a population study comparable in scope to the VLTI/MIDI sample \citep{burtscher2013, lopezgonzaga2016} which caused the paradigm shift to disk+wind models.

\part*{Conclusions} \label{sec:conc}
In this paper we present extragalactic results from open loop Fizeau imaging with LBTI/NOMIC. We observed NGC 1068 with a $22.8 \times 8.4$~m synthetic aperture at 8.7~$\mu$m; after PSF deconvolution we obtained  a measured $47\times90$~mas ($3.3\times6.2$~pc) resolution. Through the resulting images we resolved substructures of the circumnuclear dust, tested the current paradigm of disk+wind AGN models, and spatially distinguished the dusty wind from jet-host interactions. More specifically, in this work we:
\begin{enumerate}
    \item Developed the methods to deconvolve the PSF of complex, large-scale emission in LBTI/NOMIC Fizeau images,
    \item Used the correspondence between the Fizeau images and existing observations of the 5 GHz, \oiii, NIR lines, and MIR dust emission to identify different emission mechanisms and to create a self-consistent schematic of the MIR dust emission from subparsec to $\sim 100$~pc scales,
    \item Identified the NEX --a morphological feature which extends NE from the subparsec-scale MATISSE structure and traces the southern edge of the ionization cone-- which we claim is the resolved onset of AGN feedback via dusty winds and in agreement with the current disk+wind unified model paradigm,
    \item Measured the radial flux distribution of the 8.7~$\mu$m emission, identifying clear excesses beyond the radiative dust heating of the classical AGN torus model (by the accretion disk) at 20 and 45 pc from the AGN, and postulating that this excess emission is due to dust-jet interactions and shocks based on correspondence with the 5 GHz emission and NIR emission lines,
    \item Measured the optical depth of the inferred obscuring disk of the disk+wind model to be $\tau_{8.7\mu\rm{m}} = 0.41\pm 0.04$ between 20 and 30 pc radial distance from the AGN. From this we infer a lower limit on the dust density of the obscuring material to be $\rho_{\rm dust} \geq 9\times 10^{-27}~\rm{g}~\rm{cm}^{-3}$.
\end{enumerate}

These results show the interplay between jet-mode and quasar-mode feedback, as we find signatures of dust-jet interaction and dusty, radiation-pressure driven winds in the ionization cone. The disentanglement of jet emission, shocks, and dusty winds is important for the interpretation of observations with telescopes such as JWST in which the processes are blended due to lower resolution. We additionally constrain the obscuring disk in the disk+wind flavor of the Unified Model of AGN. Future observations with this mode utilizing the $L$-band and/or other narrow $N$-band filters will allow us to measure spatially resolved dust temperatures and further constrain emission processes. Finally, we have presented unprecedented ELT-scale images of an extragalactic source, developing the methods and tools necessary to make this LBTI observing mode usable for general science cases and tests for upcoming facilities.



{
}
\clearpage
\part*{Main Text Figures}
{
\centering
\includegraphics[width=0.99\textwidth]{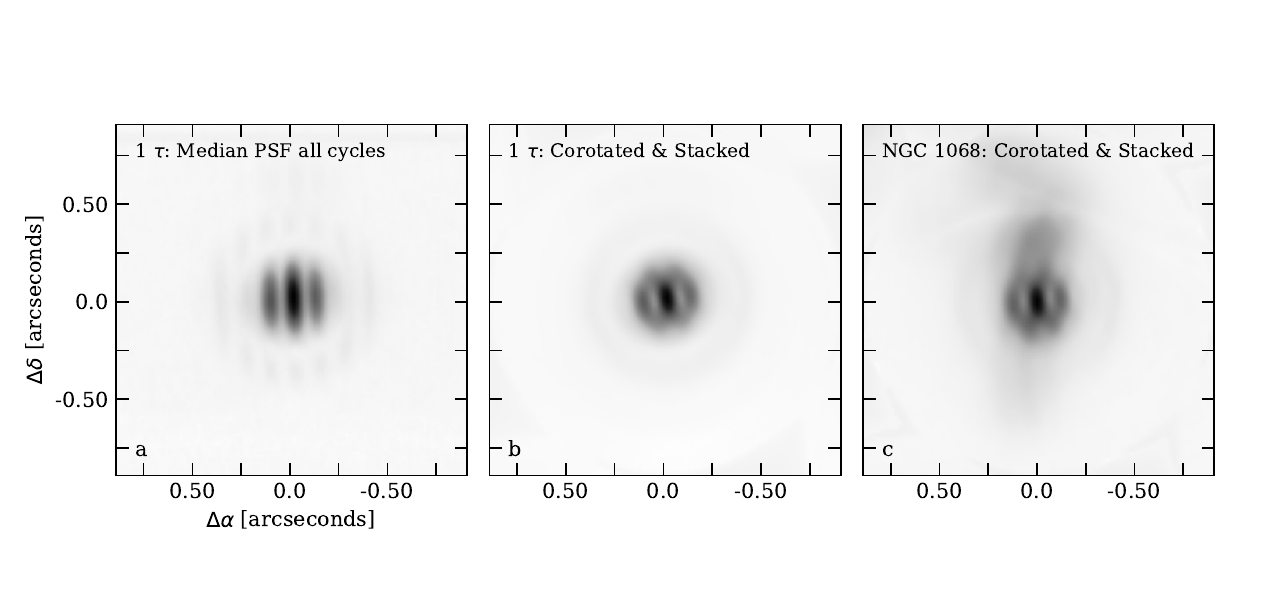}
}
\\
\noindent
\textbf{Figure 1.} Empirical PSF estimates. \textit{Panel a}: Median PSF of 1~Tau throughout the night. \textit{Panel b}: Median PSF of 1~Tau stacked and rotated to match the NGC 1068 field rotation. This serves as an estimate of the empirical PSF of the corotated and stacked science image. \textit{Panel c}: Stacked and corotated Fizeau image of NGC 1068 combining the 20 nodding cycles. In all panels, North is up and East is left.

{
\centering
\includegraphics[width=0.99\textwidth]{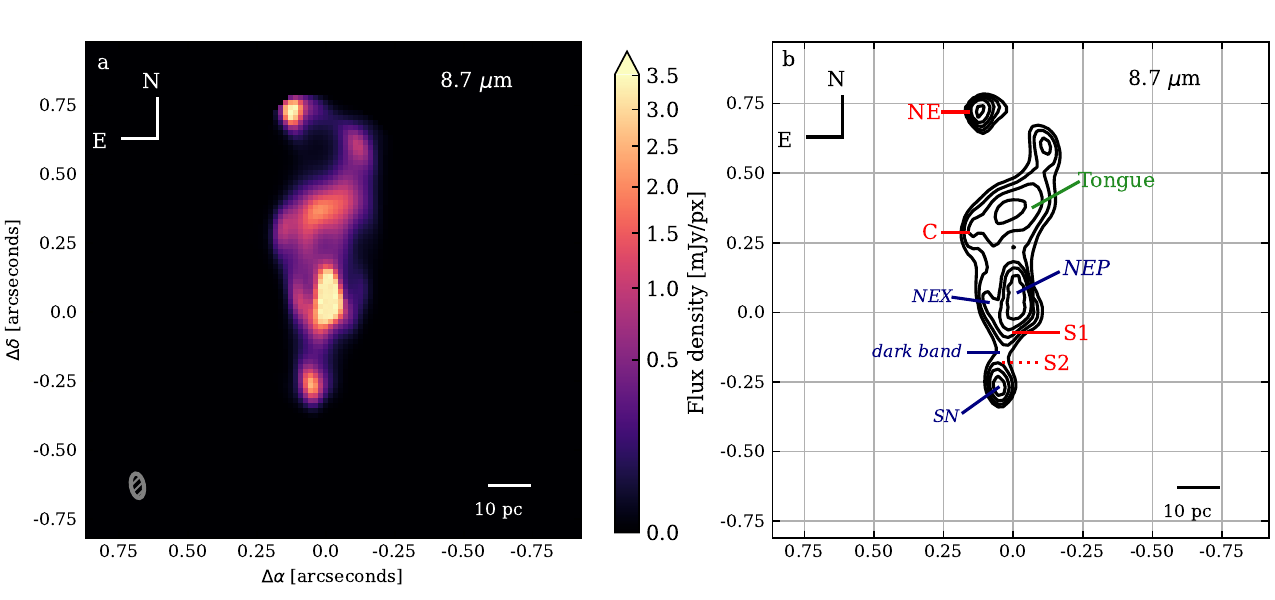}
}
\\
\noindent
\textbf{Figure 2.}
LBTI Fizeau image of NGC 1068 after deconvolution. \textit{Panel a)} Richardson-Lucy deconvolved LBTI Fizeau image of NGC 1068 at 8.7 $\mu$m using Method 1. \textit{Panel b)} Contours of the same image with feature labels. Contours start at 90\% of the peak flux density and decrease by a multiplicative factor of 2 down to a factor 128 below the maximum contour. This image was deconvolved from the stacked-corotated cycles using the Richardson-Lucy method. {The resulting resolution FWHM is $46.8\times 90$ mas ($3.3\times 6.3$ pc), shown as an ellipse in the left panel. Red and green labels indicate features from the literature (\citet{gallimore2004} and \citet{bock2000}, respectively) and blue labels indicate new features from this work: the northern extended peak (NEP), the northeast extension (NEX), the dark band, and the southern \textit{N}-band source (SN).} The results of other deconvolution approaches are given in Methods.

{
\centering
\includegraphics[width=0.99\textwidth]{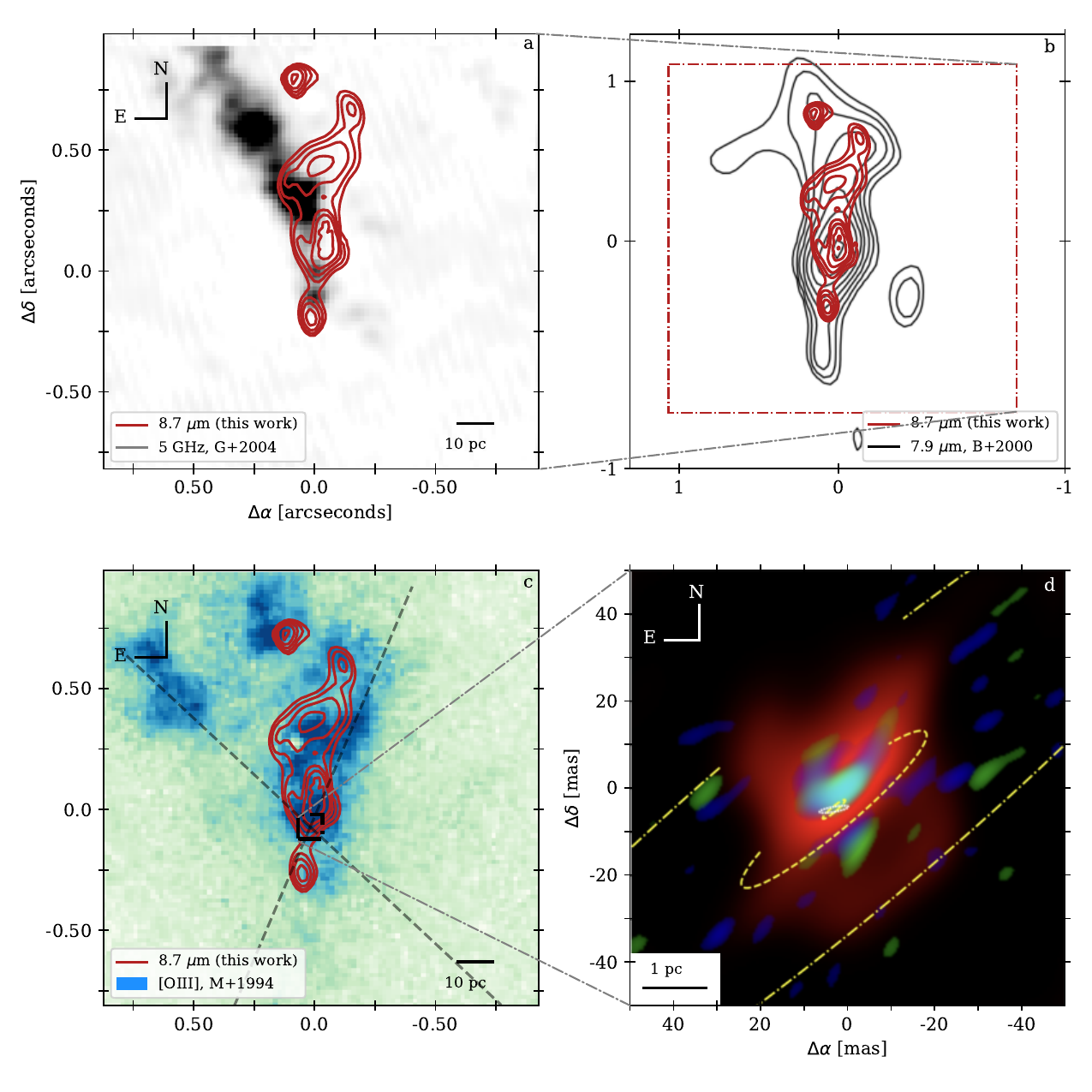}
}
\\
\noindent
\textbf{Figure 3.}
Comparison of deconvolved LBTI Fizeau image to observations at other wavelengths. Deconvolved LBTI Fizeau image of NGC 1068 at 8.7 $\mu$m compared to observations of the 5 GHz radio emission ({panel a}; \citet{gallimore2004}), the single-dish 7.9~\micron~emission ({panel b}; \citet{bock2000}), and the ionization cone traced by \oiii ~({panel c}; \citet{macchetto1994}). Panel d shows the MIR emission at sub-parsec scales as observed with VLTI/MATISSE (L-band in blue, M-band in green, and N-band in red; figure reproduced from \citep{gamezrosas2022}). The ``spur'' to the northeast of center found in the LBTI image is seen at the subparsec scale in the VLTI/MATISSE image. The yellow ellipses represent the orientation of the obscuring dust disk. The red contours are the same as in Fig. 2.


{
\centering
\includegraphics[width=0.99\textwidth]{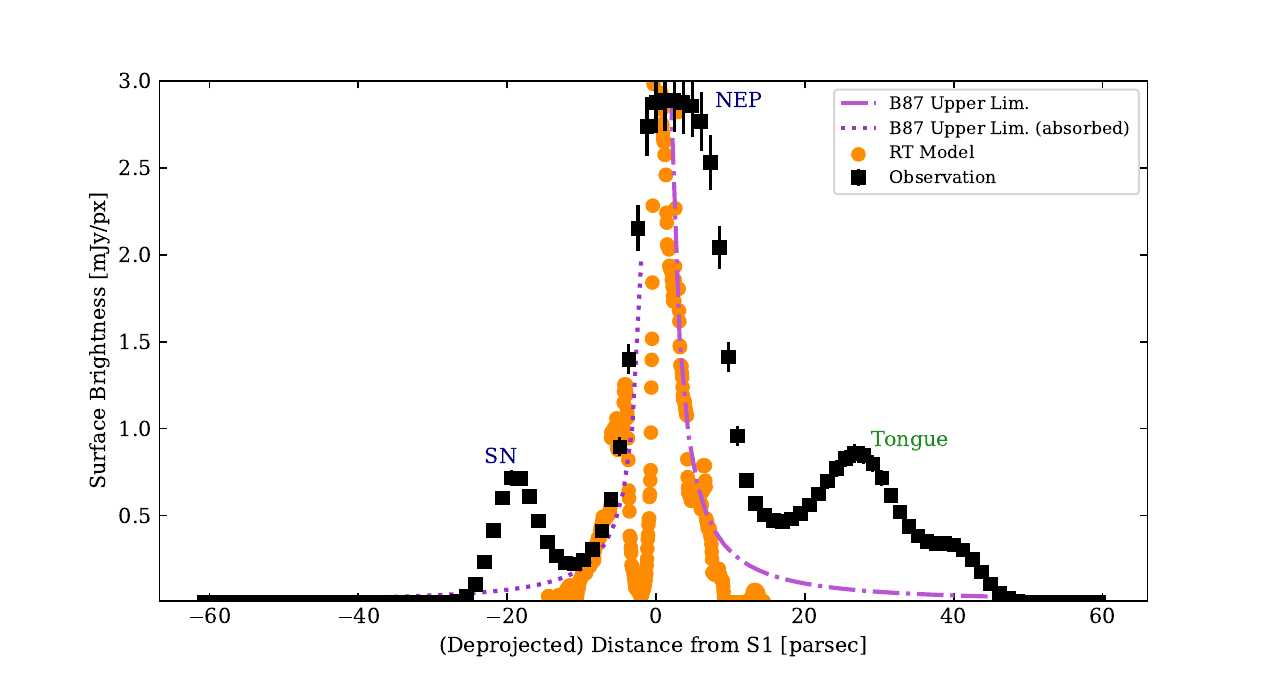}
}
\\
\noindent
\textbf{Figure 4.}
Surface brightness profiles of the observations and RT modeling. Distances have been de-projected using an inclination of $75^\circ$ from face on. Error bars on the observed surface brightness come from uncertainty in flux calibration, typically $\lesssim 10\%$.  We estimate an upper limit to the thermal dust flux as a function of radius (Methods \S4). The dashed line shows predicted upper limits, and the dotted line shows upper limits obscured by $\tau_{8.7}=0.41$. We also plot the measured flux values from a disk+wind radiation transfer model out to 15 pc. Corresponding features from Fig. 2 are labeled.


{
\centering
\includegraphics[width=0.99\textwidth]{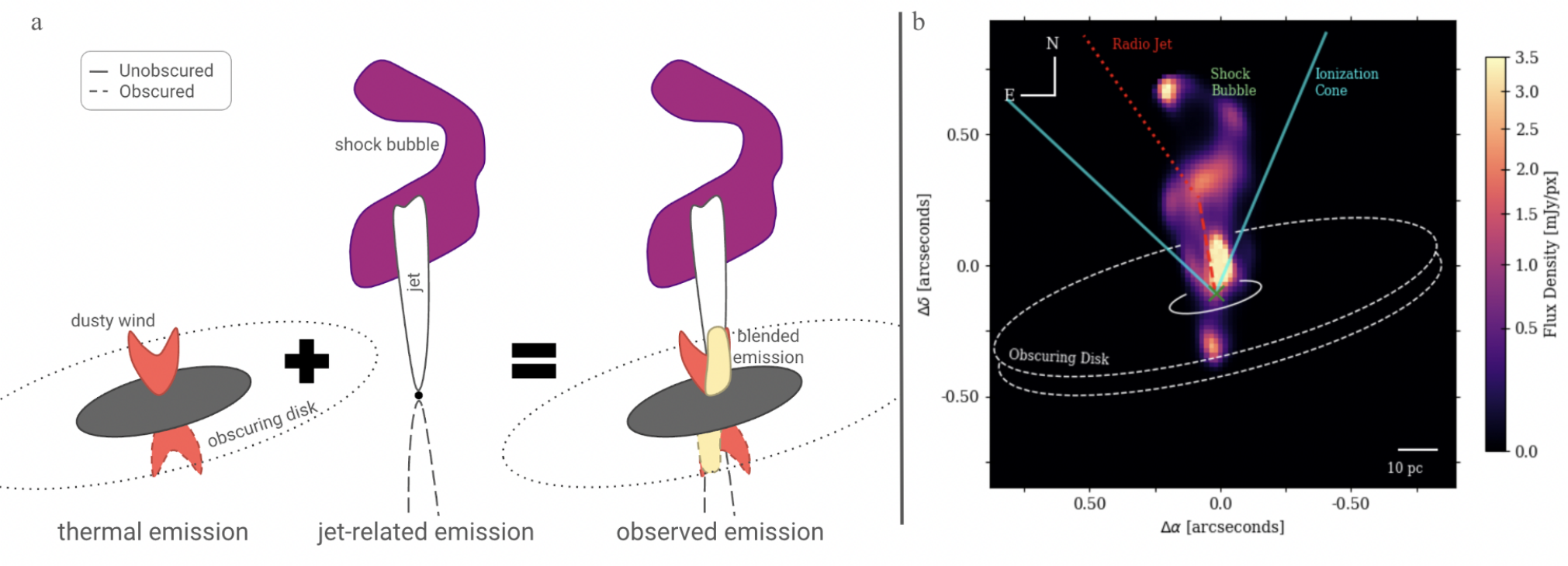}
}
\\
\noindent
\textbf{Figure 5.}
Schematic summary of the sub-parsec to $\gtrsim 100$~pc scale processes measured with the MIR emission \textit{Panel a:} A cartoon decomposition of the emission processes we identify. \textit{Panel b:} The image from Fig. 1 with labels identifying features observed in different wavelengths. The direction of the radio jet (including the bend at cloud `C') is shown in red. The edges of the ionization cone are shown in cyan. The location of the SMBH is marked with a green `$\times$' based on the `S1' position in \citep{gallimore2004}. Finally, an obscuring dust disk is indicated with white ellipses; this disk extends far enough to cover the assumed southern emission features with an optical depth of $\tau_{8.7\mu\rm{m}}=0.41\pm0.04$.



\clearpage
\part*{Methods}
\section{Observations and Calibration}
On 6 November 2023, we observed NGC 1068, 1 Tau, and Procyon using the NOMIC instrument on the LBTI in the Open-Loop Fizeau mode. We used the W08699-9\_122	filter ($\lambda_c = 8.7~\mu$m, range $[8.13, 9.35]~\mu$m) and took a number of nodding cycles on the target and calibrators consisting of 4000 exposures of 0.014 seconds (2000 exposures on target, 2000 exposures on sky). Between 05:34:42 UTC and 08:48:58 observed 20 nodding cycles toward NGC 1068.  Before and after the science target we observed 1 Tau and Procyon to serve as flux and point spread function (PSF) calibrators. 

In order to identify good and bad chop-nod cycles, we inspect the peak flux of each cycle throughout the night. Bad cycles are expected to be those with significant thermal background variations or thin clouds. We plot the peak flux over time of both NGC 1068 and 1 Tau in Methods Supplementary Fig. 1. From this we see that 1 Tau produced the same peak flux values at the beginning and end of the night, indicating a stable thermal background. We can, however, identify two cycles with low peak fluxes. Upon inspection, we find these were affected by thin cirrus clouds causing a variable background and bad adaptive optics performance. We discard these when computing statistics for the 1 Tau PSF. For both 1 Tau and NGC 1068, the cycle-to-cycle flux variations are found to be similar to the peak flux variations within a cycle. We conclude that neither the mean PSF nor the flux calibration changes throughout the night. This allows us to work with stacked images.

The ideal PSF of the LBTI is the interference pattern of two 8.4~m circular apertures separated by 14~m. The resulting 22.8~m maximum baseline gives 78.7 mas resolution (5.5 pc) at 8.7~\micron. The interference fringes exhibit a narrow central fringe peak, and a set of side lobes which rapidly decrease in amplitude as one moves further from the center. We refer to the interference fringe pattern as a Fizeau PSF. Over the span of approximately 3 hours on target, there was $59^{\circ}$ field rotation. {Field rotation causes the 22.8 m baseline of the LBTI to probe the target at different angles; after $180^{\circ}$ field rotation this approximates a single 22.8 m telescope. Simulated field rotation on a real Fizeau PSF exhibits an approximately circular, 22.8~m PSF after $\approx 60^{\circ}$ \citep{isbell2024}.} Robust estimates of both the individual cycle PSF and the PSF after field rotation are necessary for proper image deconvolution to recover the true flux distribution of the science target. The resulting corotated and stacked target and PSF calibrator are each shown in Fig. 1. 

Fringe tracking was done in the so-called ``open loop'' mode, which means that not all frames are necessarily at the zero-order (or white-light) fringe and may jump due to atmospheric shifts to higher order fringes that are not useful for this study.   However, the source was also imaged on the LMIRCam \citep{skrutskie2010} detector in the \textit{L} band {(2.8-4.1~$\mu$m)} and the Fizeau PSF was spectrally dispersed.  The interferometric phase on LMIRCam (which in Fizeau imaging has the same optical path delay as NOMIC) was set to the white-light fringe and phase shifts were corrected manually by keeping the fringes straight. While the NOMIC fringes moved due to atmospheric piston during and between exposures, this ensured that the movement was generally centered around the white-light fringe.


The Fizeau PSF may also move in the image plane or the two individual telescope PSFs may become slightly misaligned due to residual jitter from the AO system, though this effect is generally small compared to the 8.4~m N-band PSF and the precision at which the two PSFs can be overlapped ($\sim10$~mas differential tip-tilt jitter \citet{ertel2018,ertel2020}). We account for both of these effects using ``lucky Fizeau fringing" -- i.e., we discard exposures with misaligned fringes or poorly overlapped telescope beams. We first compute the 2D cross-correlation function between the theoretical PSF and each frame of our observed PSF calibration stars. In each nodding cycle, we keep the frames which have the 10\% largest cross-correlation values; in other words the 10\% of frames most similar to the theoretical PSF (a similar approach was used in \citet{leftley2021}). 
The selection of 10\% is somewhat arbitrary, but larger values tend to decrease the peak SNR of the PSF image by ``blurring'' the PSF, and smaller values have too few frames to build up a robust empirical PSF estimate. 
Within each nodding cycle, the top 10\% of frames are stacked, with each frame having been shifted to the pixel location with the highest 2D cross-correlation value, typically $\leq$ 3 pixels (pixel scale is 17.9 mas/px).

Two cycles of 1 Tau (cycles 5 and 6) showed significant deviations from the theoretical PSF, with broad, blurred artifacts. Upon closer inspection, these cycles suffered from light cirrus passing through. We therefore discard them as outliers. The remaining target and calibrator stacked PSFs show stability throughout the night. Finally, we compute the mean PSF for each calibration star. 
The mean PSF for 1 Tau is used in the following as our empirical PSF estimate because its flux (11.3~Jy; \citet{cruzalebes2019}) is more similar than Procyon's (74 Jy; \citet{cruzalebes2019}) to that of NGC 1068 ($10.2\pm3.6$~Jy; \citet{asmus2014}), and so noise estimates are similar. The empirical PSF is shown in Fig. 1, illustrating the characteristic triple-peak fringe pattern of the LBTI. 

Good frames of NGC 1068 were selected in a similar way as for the PSF calibrators. Within each nodding cycle, we compute the 2D cross-correlation function between NGC 1068 and a nominal PSF (in this case, the measured PSF from the calibrator star). The frames with the 90th-percentile largest cross-correlation values are kept. As before, the target frames are re-centered to the pixel location with the peak 2D cross-correlation value and then stacked. 

We obtain 20 stacked images of NGC 1068, one for each nodding cycle throughout the night. The resulting stacked image from each nodding cycle is shown in Methods. 10\% of frames corresponds to 200 frames per cycle; this results in exposure time of $200 \times 0.0137$~s$= 2.74$~s per cycle or $54.8$~s total exposure on NGC 1068. Before stacking each cycle, we corotated each sub-frame such that North is oriented along the y-axis in the frame. To create an estimate of the final, rotated PSF, we corotated the PSF calibrator frames by the same amount as the target sub-frames within each cycle and stacked.

We use the Mid-infrared stellar Diameters and Fluxes compilation Catalogue (MDC; \citet{cruzalebes2019}) to obtain the $N$-band fluxes of the calibrators. As is typical, we assume the unresolved stellar flux is distributed completely within the Fizeau PSF of the calibrator. This gives us a scaling between counts and Jy in each pixel that we apply to the science target. The results are consistent between Procyon (74 Jy) and 1 Tau (11.3 Jy), but we use the values from 1 Tau because it is closer to the nuclear flux of NGC 1068 ($10.2\pm3.6$~Jy; \citet{asmus2014}) and thus has a similar signal-to-noise ratio.  

{
As LBTI Fizeau imaging is rather new to the community, we summarize the main differences from traditional, single aperture observations. 
First, the Fizeau PSF is characterized by a ``triple peaked'' fringe pattern, yielding 22.8~m resolution in one direction and 8.4~m resolution in the orthogonal direction. We show that even small amounts of field rotation allows one to obtain resolution better than 8.4~m in all directions, though a circular PSF would require $180^{\circ}$ rotation \citep{isbell2024}. Second, for non-ideal field rotation (and generally for the non-circular PSF of any segmented primary mirror), the effects of the resulting elongated PSF can be removed using image deconvolution. While image deconvolution is rather commonplace for observations with e.g., Keck\citep{bock2000} or JWST\citep{leist2024}, we explore various methods below for the specific LBTI configuration. Third, we employed ``lucky fringing,'' keeping only the best 10\% of data. This extremely conservative approach was feasible due to the brightness of the source, but recently published LBTI results have shown that $\geq 90$\% of frames can be kept in some cases for Fizeau imaging\citep{isbell2024}, greatly increasing observation efficiency. For sources with $K < 4.5$~mag, fringe tracked imaging can be performed, removing the need for frame selection. A scheduled upgrade to the fringe tracking detector will increase this limit to $K < 10$~mag\citep{conrad2023}. Finally, because the Fizeau aperture consists of two 8.4~m telescopes, the collecting area equivalent to that of a $11.9$~m telescope (111~m$^2$). Upcoming facilities, such as the ELT and GMT, will have larger effective collecting areas (978~m$^2$ and 368~m$^2$, respectively) and thus higher sensitivity than the LBTI. 
}

\section{Image Deconvolution}
\label{sec:deconv}
As can be clearly seen in Fig. 1, the stacked and corotated science cycles exhibit significant artifacts due to the Fizeau PSF. While extended flux features are still discernible, small-scale features near the center of the image are dominated by these artifacts. Using empirical measurements of the Fizeau PSF in individual cycles and for the stacked image, we can deconvolve the known PSF features to reveal the underlying source flux. This was previously done in \citep{leisenring2014} for observations of Io, but in the intervening years there have been changes to the adaptive optics performance of the LBTI\citep{pinna2016} and we used the NOMIC detector instead of LMIRCam. We therefore attempt Fizeau PSF deconvolution in three different ways: the first two use Richardson-Lucy deconvolution as in \citep{leisenring2014} and the third was developed for this work based on the H\"ogbom CLEAN algorithm \citep{hogbom1974}.

\subsection{Richardson-Lucy Deconvolution}
We proceed two different ways using Richardson-Lucy deconvolution \citep{richardson1972, lucy1974}:
\begin{enumerate}
    \item Stack all corotated frames and deconvolve with a stacked, corotated PSF estimate, or
    \item Deconvolve the image from each cycle with the PSF calibrator and then stack the resulting images.
\end{enumerate}
The first method has the benefit of doing as few convolutions as possible (each introduces noise) at the expense of requiring that the estimate of the final PSF is robust. The second method uses a robust, single-cycle PSF but requires 20 (lower SNR) deconvolutions instead of one. 
In both cases, we use Richardson-Lucy deconvolution (from \pack{scikit-image.restore}) because of its robustness to noise and efficiency. We find that both sequences give similar results, which is strong indication of the validity and robustness of the target flux distribution we recover. In practice, stacking before deconvolution (Method 1) increased the SNR of faint features, as noise is added destructively. This led to more stable results when changing the parameters \code{n\_iter} and \code{filter\_epsilon} (the number of iterations and the clipping value for avoiding division by zero, respectively).

We estimate the uncertainty of each pixel in the resulting images using a bootstrap method; we perturb the empirical PSF by Gaussian noise and re-performed the deconvolution numerous time. For the images presented in Main we used Method 1 with \code{n\_iter=32} and \code{filter\_epsilon=0.005}; we show the results in Fig. 2.  The total flux of the image was not conserved during deconvolution, and therefore the deconvolved image is re-normalized so that the total flux of the resulting image equals the total flux of the corotated+stacked image (specifically, it is scaled by a factor of 0.7). The resulting PSF is $46.8 \times 90$~mas FWHM.

As expected from previous observations, we see significant extended flux along the north-south direction as well as a bright point source to the northeast at 600 mas (45 pc). However, the 78 mas PSF from the 22.8~m baseline allows us to resolve inner substructure near the AGN core. Here we see a ``spur'' of flux pointing toward the northeast out to $\sim 10$~pc. In the deconvolved-then-stacked (Method 2) image, we see evidence of an additional, smaller spur to the northwest. These spurs are found near the expected AGN location and are oriented in a similar direction as extended $N$-band dust measured on smaller scales with MATISSE \citep{gamezrosas2022}. In Method 2, we find that the bad channel (visible in Supplementary Fig. 2 in the upper third of the images) causes more significant artifacts than in Method 1. In ED Fig. 1, we see a negative contour at the location of the bad channel when using Method 2. It seems that the stacking of images in Method 1 washes out the effects of this channel. Due to the stability of the PSF throughout the night which allows for robust image stacking and the strong artifacts present in Method 2, we use Method 1 for analysis throughout this article. We note, however, that results only change by a few percent when using the other deconvolution method.


\subsection{CLEAN-inspired method}
The H\"ogbom CLEAN algorithm is an aperture synthesis algorithm, developed to deconvolve a known ``beam'' from a measured radio emission map affected by prominent and extended sidelobes \citep{hogbom1974}. Similarly, the known ideal Fizeau PSF produced by the LBTI has prominent and extended sidelobes. Due to the high fidelity of CLEANed radio images, we developed an experimental implementation for LBTI Fizeau images which iteratively removes the known, extended PSF. 

We use the following notation:  \texttt{Im$_0$} is the initial, “dirty” image to deconvolve, \texttt{beam} is the known PSF estimate, \texttt{gain} is scaling applied to the PSF to remove flux from the dirty image in each iteration, \texttt{CUTOFF} is the flux level to which the image is cleaned, \texttt{MAXITER} is the maximum number of iterations to compute, and \texttt{peak} is the maximum flux in the current image, \texttt{I$_i$}. The algorithm then proceeds as follows:
{\texttt{
\\START\\
WHILE peak > CUTOFF AND $n_{\rm iter}$ < MAXITER
\begin{enumerate}
\item Find peak flux in image, at coordinates ($x_p,y_p$)
\item Create Im$_{i+1}$ 
\begin{itemize}
    \item Create a zero-image with same dimensions as Im$_0$
    \item Place a point source with flux = gain $\times$ peak at ($x_p,y_p$) and convolve with beam
    \item Subtract this from Im$_i$ to get Im$_{i+1}$
    \end{itemize}
\item Put a point source at ($x_p,y_p$) in the CLEAN image
\item $n_{\rm iter} = n_{\rm iter} + 1$
\end{enumerate}
ENDWHILE\\
END
}}

We applied the above method with \code{gain = $10^{-4}$}~and \code{MAXITER = $10^6$} on each of the 20 exposure cycles. We also set \code{CUTOFF = Mean(Im$_0$)}, i.e., stop when the iteration peak is just above the background level. We finally corotate and stack the images as above. The resulting, deconvolved image is shown in ED Fig. 2. The results of this method show more low-surface brightness features than the Richardson-Lucy deconvolution in Fig. 2; in particular, the flux extending to the south of the photocenter and the flux within the northernmost arc are more prominent in the CLEAN-inspired deconvolution. Importantly, the ``spur'' extending to the NE of the photocenter appears in this deconvolution as well, with a similar opening angle and length. The fact that this feature appears in this independent deconvolution method gives yet more credence to its fidelity. 

\subsection{Neural Network Deconvolution}

The method presented here is a new algorithm developed to perform image deconvolution without using an experimental PSF. Instead, we use a neural network trained on a large dataset of simulated LBTI observations. This method is derived from the interferometric image reconstruction algorithm proposed in \citep{Sanchez2022}. The results are presented in ED Fig. 3.

The algorithm steps are as follows:
\begin{enumerate}
    \item Generate a large set of mock astrophysical images.
    \item Simulate random LBTI PSFs, convolve the images with these PSFs, and introduce noise.
    \item Train a neural network to restore the original images using only the simulated data.
    \item Apply the trained neural network to the actual astronomical observation.
\end{enumerate}

We elaborate on these steps in the subsequent paragraphs.

\bigbreak
\textbf{I. Generation of mock astrophysical images}
\bigbreak

The objective is to rapidly produce a diverse set of images containing varied structures with different spatial scales. We create each image by summing several component images, each representing structures at a characteristic spatial scale, $\lambda$, randomly selected between 1 pixel and the full field of view (FoV). The number of component images, N, varies randomly from 1 to 20. For each component we:
\begin{enumerate}
    \item Compute a 2D discrete exponential power spectrum P with characteristic scale $\lambda$.
    \item Assign a random phase $\phi$ to each pixel.
    \item Calculate the real part of the Fourier Transform of $P \times e^{i \times \phi}$, and mask the negative values.
\end{enumerate}

The N images are combined using random weights, and the total flux of the resultant image is normalized to unity. This process is repeated to generate $2^{20}$ unique images of dimensions ($n_{pix} \times n_{pix}$), with $n_{pix}=100$.

\bigbreak
\textbf{II. Simulation of Observations}
\bigbreak

Each generated image is then used to simulate an observation cube with dimension ($n_{frames}, n_{pix}, n_{pix}$), with $n_{frames}=32$. The decision to use data cubes with $n_{frames}$ rather than a single frame is driven by two factors: first, this approach allows the neural network to prioritize data from the highest quality frames, thereby optimizing image selection; second, it enables differentiation of PSF variations—which change from frame to frame—from static astrophysical elements.

Instantaneous PSFs are computed as the Fourier Transform of $Pup \times \exp(i \times \phi_{res})$, with $Pup$ the telescope pupil and $\phi_{res}$ the residual atmospheric phase post-adaptive optics correction. The pupil is modeled as two 8.4 m disks, each with a central 0.9 m obstruction, and a 14.4 m center-to-center separation. $\phi_{res}$ is calculated from a sum of Zernike polynomials with coefficients randomly selected following a power-law distribution $P(n) \propto n^{-\alpha}$, with $n$ the radial order of the polynomial. 

A number $n_{\tau}$ of these instantaneous PSFs, drawn uniformly between 1 and 100, are averaged to create a semi-long exposure PSF. Each image is then convolved with $n_{frames}$ such PSFs, creating an observation data cube.

Two types of normal noise are added to the datacube, one proportional to the square root of the surface brightness of the images (photon noise of the source), and one uniform across the FoV (photon noise of the subtracted sky). The overall noise level is randomly drawn for each observation to produce a SNR between 1 and 50 for individual frames.

\bigbreak
\textbf{III. Neural Network and training}
\bigbreak

After generating the training dataset, consisting of $2^{20}$ random images and the corresponding observations, we train a neural network to reconstruct the original images from these convolved, simulated datasets. 

We use a simplified version of the U-NET architecture, which is fully convolutional and includes skip layers. A schematic is shown in Supplementary Fig. 4.

The training is done on 50 epochs, each one including the entire dataset in batches of 16 observations. We use the Adam optimizer with a mean absolute loss function and an initial learning rate of $5\times^{-4}$. This learning rate is halved each time there is no improvement in the loss after five consecutive epochs. The final mean absolute error, indicating the difference between the network's prediction and the actual original images, is 0.09 for both the validation and the training dataset, demonstrating that the training did not result in overfitting.

\bigbreak
\textbf{IV. Application to the actual observations}
\bigbreak

The observations consist of 20 cycles, corresponding to 20 different orientations of the source in the plane of the sky, each of these cycles corresponding to 2000 individual short-exposure frames.

We split each of these 2000 frame sets into 62 data cubes each with 32 frames, which were fed to the trained neural network, hence producing 62 deconvolved images for each orientation. These 62 deconvolutions are then averaged to produce one image per orientation. The 20 resulting images are then de-rotated, and the final image is computed as the median of these images. It is shown in ED Fig. 3. The low surface-brightness features are better recovered than in Fig. 2, and resemble those of the CLEAN approach in ED Fig. 2. The central emission is more compact than the CLEAN deconvolution results, but shows a similar NW-SE extension. The NEX feature, however, is not as readily apparent in these results, though the flux envelope near the center traces roughly the opening angle of the ionization cone. 

\subsection{Estimating Image Significance}
The standard deviation of the cycle PSFs is used as an uncertainty estimate, allowing us to bootstrap errors of the deconvolution we perform on the target frames. In Supplementary Fig. 5 we show all 1 Tau cycles' PSF images and flux profiles. We compute the per pixel mean and standard deviation of all cycles (excluding cycles 5 and 6 due to thin cirrus). In 1000 bootstrap iterations, we perturb the empirical PSF (Fig. 1) with random noise set by the standard deviation at each pixel. We then stack and corotate the perturbed PSF to match the rotations of the 20 target observing cycles. From the 1000 iterations, we can compute the uncertainty per pixel for the recovered deconvolved images below; from this we conclude that all primary structures in the image (the central component with a ``spur'', the southern point source, the northern extended structure, and the northernmost point source) are robustly recovered at $\gtrsim 10\sigma_{\rm bootstrap, px}$. As an example, we display the signal-to-noise ratio (SNR) map for deconvolution Method 1 with 8.7~$\mu$m image contours overlaid in Supplementary Fig. 6; this shows that all primary features of the image are found at SNR $> 10$.

\section{(Non)thermal Dust Emission Along the Radio Jet}
With apertures having a diameter of 8 pixels (0.144 arcsec, 10.1 pc), we measure the flux at several image locations. We measure the flux of the southern component (SN) to be $F_{\rm South} = 39.6 \pm 2.4$~mJy, with the aperture centered 25.1 pc to the south of the center. At the same distance to the north (in cloud `C'), we measure the flux to be $F_{\rm North} = 100.6 \pm 6.2$~mJy.

Utilizing Eq. 1 from \citep{barvainis1987}, we can estimate dust temperature as a function of radius:
\begin{equation}
    T_{gr}(r) = 1650 \Big(\frac{L_{\rm acc}}{r^2} \frac{{\rm pc}^2}{10^{10}{\rm L}_{\odot}} \Big)^{1/5.6} e^{-\tau_{\rm uv}/5.6} {\rm K},
    \label{eq:barvainis}
\end{equation}
where $L_{acc}$ is the luminosity of the accretion disk in $L_{\odot}$, $r$ is the distance from the accretion disk in parsec, and $\tau_{\rm uv}$ is the optical depth to the ultraviolet continuum. The luminosity of grains of this temperature is then $L_{\nu,\Omega}(r) = \sigma/ \nu\pi \times T_{gr}(r)^4  ~{\rm W}~{\rm m}^{-2}~{\rm Hz}^{-1}~{\rm sr}^{-1}$. The predicted thermal dust flux in an aperture is $F_{\rm 8.7{\mu}m, therm} = L_{\nu, \Omega}(r) A$, where $A$ is the aperture area. We show in Fig. 4 that there are significant flux excesses 20 and 45 pc from the core which are associated with 5 GHz features `S2,' `C,' and `NE.'

As stated in the Main Text, the contribution from non-thermal sources within cloud C is very small in the MIR. 
The authors in \citep{villarmartin2001} present several possible dust-heating mechanisms within shocks for NGC 1068, but there has been some disagreement in the literature about the prevalence of shocks along the radio jet in NGC 1068. The authors of \citep{kraemer1998} claim that shocks are necessary to heat the dense clouds at their ``position 2," (cloud C in \citet{gallimore2004}) but \citep{martins2010} claim photoionization actually could be sufficient with different assumptions of cloud densities. More recently, \citep{kakkad2018} show high cloud column densities, indicating that shocks are indeed required. 

The MIR emission of NGC 1068 in the central 50 pc seemingly requires contributions from several distinct thermal and non-thermal processes. Future LBTI observations of this and other AGN, either with a number of $L$-, $M$-, or $N$-band filters to get dust color temperatures or with narrow filters targeting emission lines (e.g., [S IV]~10.5~$\mu$m and [Ne II]~12.8~$\mu$m), will be able to distinguish the emission or heating mechanisms. 

\section{Radiation Transfer Modeling}
\label{sec:rt}

The models which most closely match interferometric data of nearby Seyferts (e.g., \citep{stalevski2019,leftley2023}) consist of a compact, dusty disk and a hollow hyperbolic cone extending in the polar direction. Here we modify the best-fitting disk+wind model of \citep{stalevski2019} to use the luminosity of NGC 1068 ($10^{12}$~L$_{\odot}$) and the opening angle and inclination fit by \citep{leftley2023}. We then increase the box size of the pre-existing small-scale circumnuclear model to measure the flux at $15$~pc scale. Except when otherwise listed, the parameters of the disk+wind model are the same as used in \citep{isbell2022}. We use $i=75^{\circ}$, a hyperbolic cone opening angle $\theta_{\rm open} = 45^{\circ}$, an outer extent of the hyperbolic cone of $r_{\rm hyp,out}=24$~pc, a disk optical depth of $\tau_{\rm disk, 9.7} = 15$ (the same as in \citep{isbell2022}), and a disk outer radial extent of $r_{\rm disk,out}=10$~pc. We also set the instrument field of view to 30 pc and the number of pixels in each direction to 600 pixels. The resulting model is shown in 
ED Fig. 4, 
and we also extract radial flux profiles to compare to our measured emission profiles.

The main goal of the RT modeling was to determine whether the NEX in our images could realistically be centrally-heated dust in a windy outflow. Morphologically, there is indeed a similar structure due to edge-brightening of the wind. To test whether the flux was plausible we placed an aperture (3.8 pc radius) at approximately the same location in both the Fizeau image and the RT model. The aperture was placed on the southern edge of the ionization cone in both cases, 7 pc from the nucleus. The RT model aperture contains $\sim 100$~mJy of flux with a peak flux density of 1.2 mJy/px. The observation aperture contains $49 \pm 3$~mJy with a peak flux density of 1.5 mJy/px. While a detailed comparison of the resulting fluxes is not meaningful, the disk+wind RT model shows that it is plausible that the emission in the NEX is AGN heated dust at the edge of the outflow cone. A detailed RT model parameter study, comparing to the combined VLTI/GRAVITY, VLTI/MATISSE, and LBTI data, will be the subject of future work. 

\section{Optical Depth}
The dusty wind, ionization cone, and radio jet are all presumed to be symmetric above and below the accretion disk. Based on correspondence with the radio data, we place the SMBH at the location marked with a green `$\times$' in Fig. 5, near the bottom of the brightest flux region. If we then assume that similar structure should extend to the north and south, it is immediately apparent that some absorbing medium is present -- likely the disk-like component of the dusty torus.

The optical depth is given by
\begin{equation}
    \tau = - \ln \frac{F_{\rm South}}{F_{\rm North}} = lN\sigma = l \kappa \rho,
    \label{eq:tau}
\end{equation}
where $F_{\rm South}$ and $F_{\rm North}$ are the fluxes equidistant south and north of the SMBH, $N$ is the number density of the obscuring material, $\sigma$ is the material's cross-section, $\kappa$ is the mass extinction coefficient of the material, $l$ is the path length of the absorbing material, and $\rho$ is the density of the material. Assuming a specific dust composition, we are left with two free parameters representing the density of the dust and the path length (i.e., the thickness of the obscuring structure). Taking the measurements from the same apertures as above (Methods \S4), we find that the southern emitting region would need to be obscured by an object with an optical depth of $\tau_{\rm 8.7} = 0.41 \pm 0.04$.

The optical depth of such a structure should increase as one gets closer to the SMBH. We measure the average value at 25.1 pc within a 10.1 pc diameter aperture. Previous measurements \citep{gamezrosas2022} at $\lesssim 1$~pc give $\tau_{8.7\mu\rm{m}} = 0.94-1.51$ (their measurement is at a slightly longer wavelength, so we scaled their reported optical depth by a factor 0.74 to compare at 8.7~$\mu$m, based on the extinction curve from \citet{schartmann2005}). Assuming a linear change in $\tau$, we obtain an optical depth gradient of $-0.05~\rm{pc}^{-1} \leq \Delta\tau_{8.7\mu\rm{m}} \leq -0.02~\rm{pc}^{-1}$. 

{Since the optical depth $\tau$ is related to dust density, composition, and path length through the obscuring material, we explore what a gradient might imply. The dust composition is likely approximately constant, so we focus on dust density and path length. Current disk+wind RT models should naively lead to an \textit{increased} optical depth at larger radii because the constant-density disk becomes thicker with an opening angle of $5^{\circ}$, leading to a longer path length. We observe, however, the opposite trend. The RT modeling in this work assumes a constant dust density in the disk (based on comparisons to Circinus on much smaller scales \citet{stalevski2019,isbell2023}). Our new measurements imply the necessity of a decreasing density gradient counteracting the increase in path length. An exploration of RT parameters would be greatly aided by more constraints on dust temperatures and emission mechanisms, so this will be explored in future work after LBTI observations with more filters have been obtained.}

Rearranging Eq. \ref{eq:tau}, we find the relation $\rho = \tau_{8.7~\mu\rm{m}} / (l \kappa_{8.7~\mu{\rm m}})$. Because NGC 1068 is inclined $\approx 75^{\circ}$ from face-on, we can compute an upper limit on the path length ($l$) corresponding to a lower limit on the dust density: 
    $\rho_{\rm lower} = \tau_{8.7~\mu\rm{m}}~{\sin i} / ({d~\kappa_{8.7~\mu\rm{m}}})$.
For $i = 75^{\circ}$, $\kappa_{8.7~\mu\rm{m}} = 12116.5~\rm{m}^{2}~\rm{kg}^{-1}$ from \citep{schartmann2005}, $\tau_{8.7~\mu\rm{m}} = 0.41\pm0.04$ as above, and $d\approx 25$~pc as measured from the image, we obtain $\rho_{\rm dust} \geq 9\times 10^{-27}~\rm{g}~\rm{cm}^{-3}$. The ISM has a mean gas density of 1 atom per cubic centimeter, or $\bar{\rho}_{\rm ISM} = 1.6\times10^{-24}~{\rm g}/{\rm cm}^-3$ \citep{draine2011}. This corresponds to a dust-to-gas ratio 25~pc from the AGN of $6 \times 10^{-3}$. This is 0.6 times the mean ISM dust-gas-ratio of $\approx 1\%$\citep{bohlin1978}. A study of 36 X-ray observed AGN yielded dust-to-gas ratios in the range [0.01-1] times the standard ISM value\citep{esparza2021}; our measurement falls well within this range.

\textbf{Data availability}: Raw and processed data corresponding to the current study are available from the corresponding author upon request.

\textbf{Code availability}: The algorithms used for data processing are described in this work, but data processing Python scripts and data processing instructions are available within the public github repository located at \href{https://github.com/jwisbell/lbti_fizeau}{https://github.com/jwisbell/lbti\_fizeau}. 

\textbf{Acknowledgments:} We thank the anonymous referees for their feedback which greatly improved this work. VFG acknowledges funding from the National Aeronautics and Space Administration through the Exoplanet Research Program under Grant No. 80NSSC21K0394 (PI: S. Ertel). RGP, NM, JL, and PV been supported by the French Agence Nationale de la Recherche (ANR) through the grant ``AGN MELBa” ANR-21-CE31-0011.

\textbf{Author contributions:} JWI, J-UP, GW, MS, JLef, WJ, RP, NM, and VGR contributed to scientific analysis, modeling, and interpretation. JWI and PV developed and tested deconvolution methods. JWI, SE, AB, JC, VFG, WFH, JLei, JP, and KW took part in observations setup, execution, and processing.

\textbf{Competing interests:} The authors declare that there are no competing interests.
\clearpage
\part*{Extended Data and Supplementary Information}
{
\centering
\includegraphics[width=0.95\linewidth]{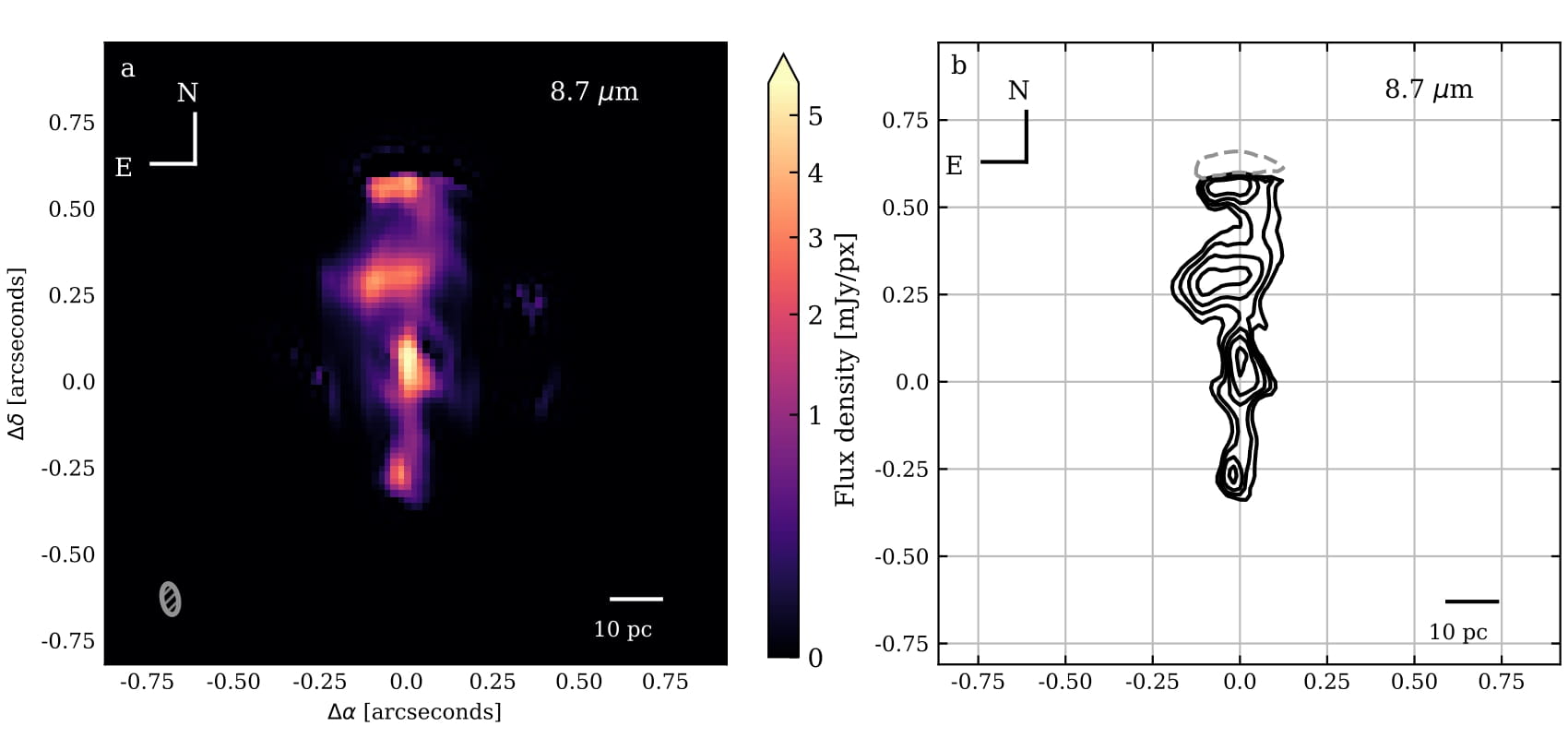}
}
\\
\noindent
\textbf{Extended Data Figure 1.} LBTI Fizeau image of NGC 1068 after deconvolution using Method 2. \textit{Panel a)} Deconvolved LBTI Fizeau image of NGC 1068 at 8.7 $\mu$m \textit{Panel b)} The [2,3,4,5,10,20,50]$\times \sigma$ contours of the same image. Method 2 used Richardson-Lucy on each observing cycle and then stacking. The results are overall similar to Method 1 (Fig. 2), but with slightly narrower features and a negative flux contour to the north due to a bad pixel channel (visible in Supplementary Figure 2).

{
\centering
\includegraphics[width=0.95\linewidth]{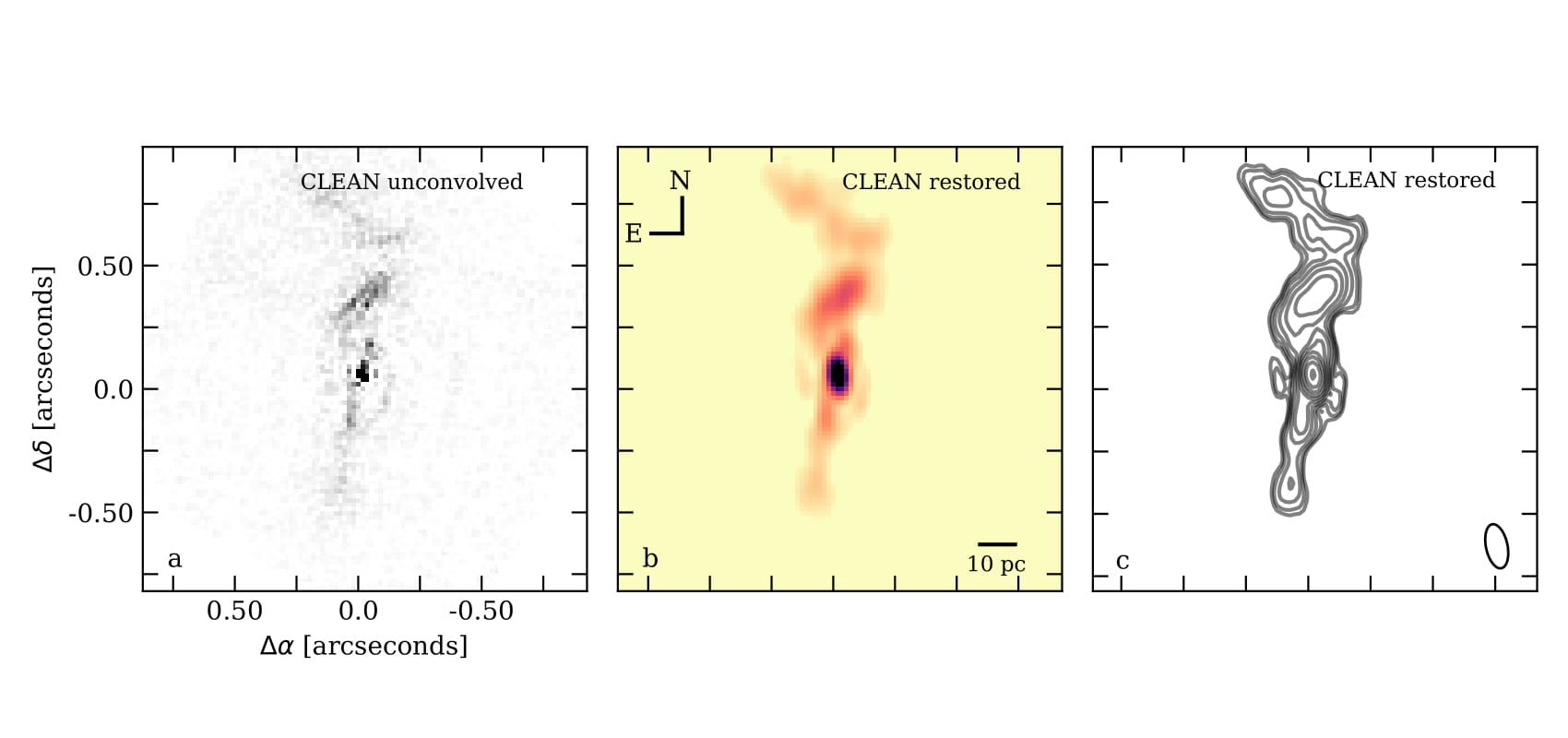}
}
\\
\noindent
\textbf{Extended Data Figure 2.} Results of the CLEAN-inspired deconvolution algorithm. \textit{Panel a)} the point source locations produced by the CLEANing iterations. \textit{Panel b)} the CLEAN-inspired image convolved with the restoring beam.  \textit{{Panel c})} the [8, 16, 32, 64]$\times \sigma$ contours of the restored CLEAN-image, with $\sigma$ measured from the image background. The FWHM of the restoring beam is given in the bottom left corner. Resulting structures resemble those of Fig. 2, albeit with more prominent low-surface-brightness features.

{
\centering
\includegraphics[width=0.95\linewidth]{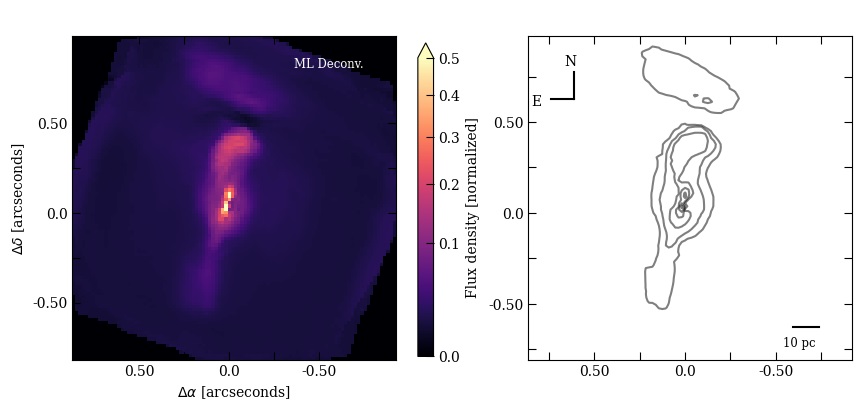}
}
\\
\noindent
\textbf{Extended Data Figure 3.} Stacked and corotated results of the neural network deconvolution. \textit{Left)} surface brightness of the recovered emission, normalized to the peak flux. \textit{Right)} Contours of the surface brightness, beginning from 90\% of the peak and decreasing by factors of 2 down to a factor 128 below the maximum contour. Low surface brightness features similar to those in Extended Data Fig. 2 are recovered, though the central emission appears to be slightly less extended.

{
\centering
\includegraphics[width=0.95\linewidth]{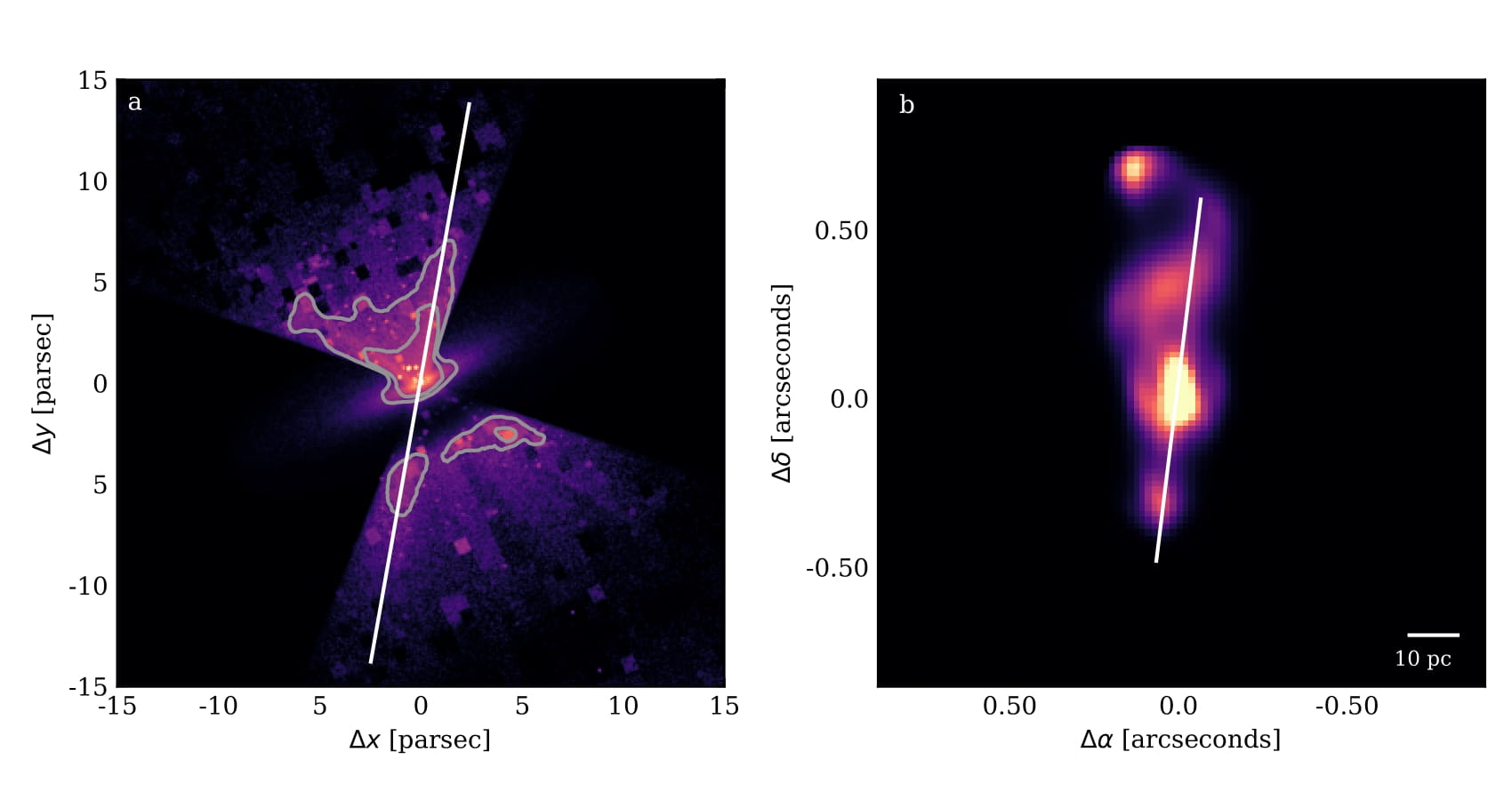}
}
\\
\noindent
\textbf{Extended Data Figure 4.} Comparison between radiation transfer modeling results and the LBTI observation.  \textit{Left)} Radiation transfer of a disk+wind model. Contours are set at $4, 8\times 10^{-10}~\rm{W~m}^{-2}\rm{sr}^{-1}$ to emphasize the edge-brightening effect and to show that a similar structures to the NEX and southernmost MIR flux we observe are reproduced. \textit{Right)} The same as Fig. 2. In both panels, the white lines represent the extraction aperture used to produce the surface brightness profile in Fig. 4.

\begin{center}
    \includegraphics[width=0.5\textwidth]{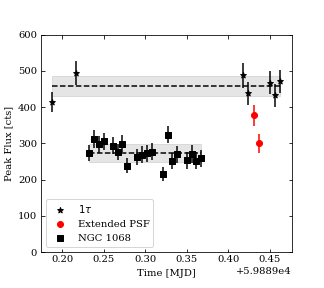}
\end{center}
\textbf{Supplementary Figure 1.} Transfer function throughout the night. The measured flux of both the calibrator and the target are shown to vary little throughout the night. Each point is the peak flux of the fringe pattern, and error-bars represent the standard deviation of the flux within each frame. They are comparable to the statistical flux variations shown over the night. The two points in red were affected by thin cirrus clouds and are discarded.

%
\clearpage
\begin{center}
\includegraphics[width=0.91\textwidth]{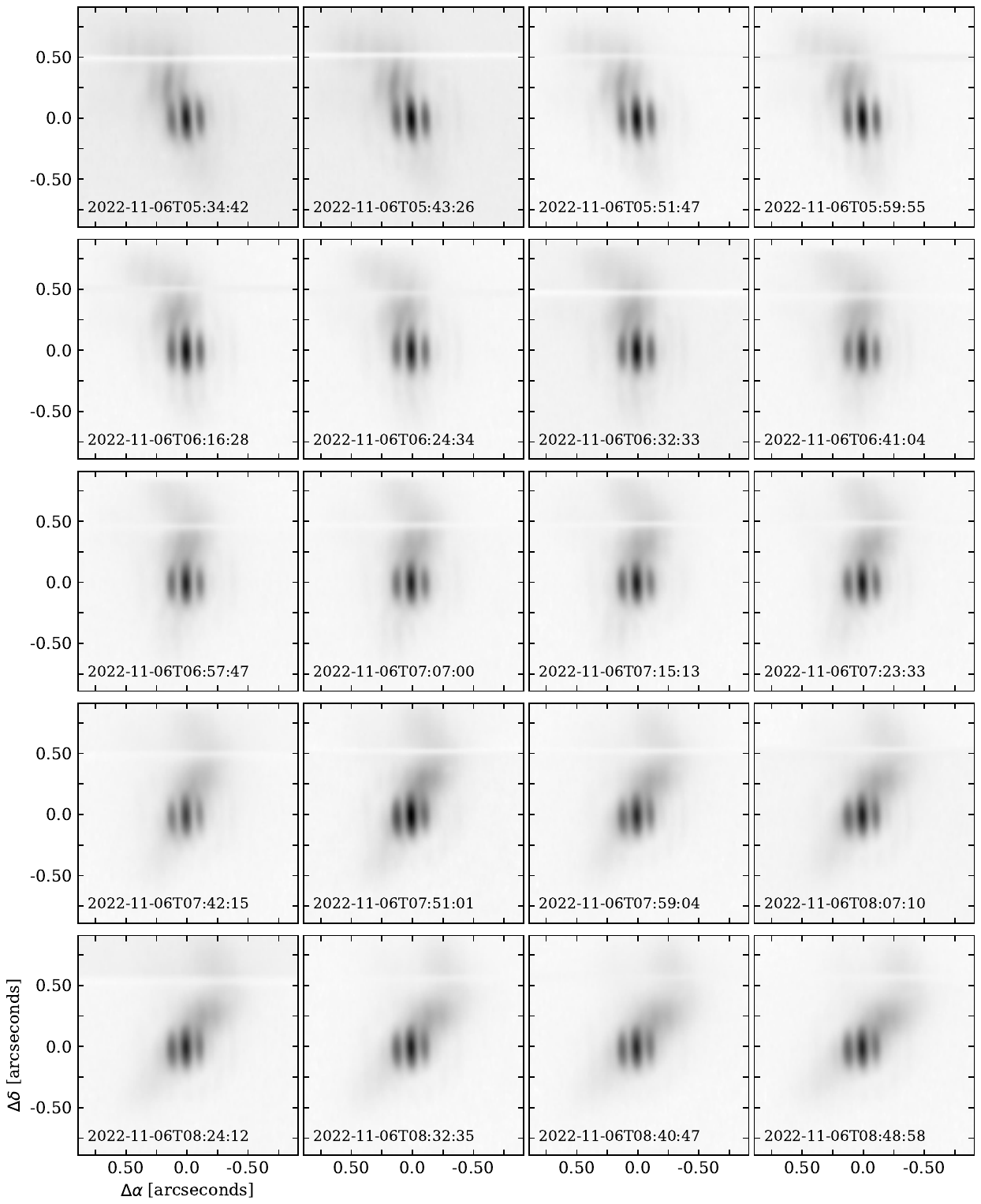}
\end{center}
\textbf{Supplementary Figure 2.} Individual exposure cycles of NGC 1068 after frame selection. The selected frames are stacked within each cycle. The summed exposure time is approximately 2.75 s for each cycle, resulting in a fringe peak SNR of $\approx 320$ in each stacked cycle image. The effects of field rotation are not corrected in these panels.
\clearpage

\begin{center}
\includegraphics[width=0.99\textwidth]{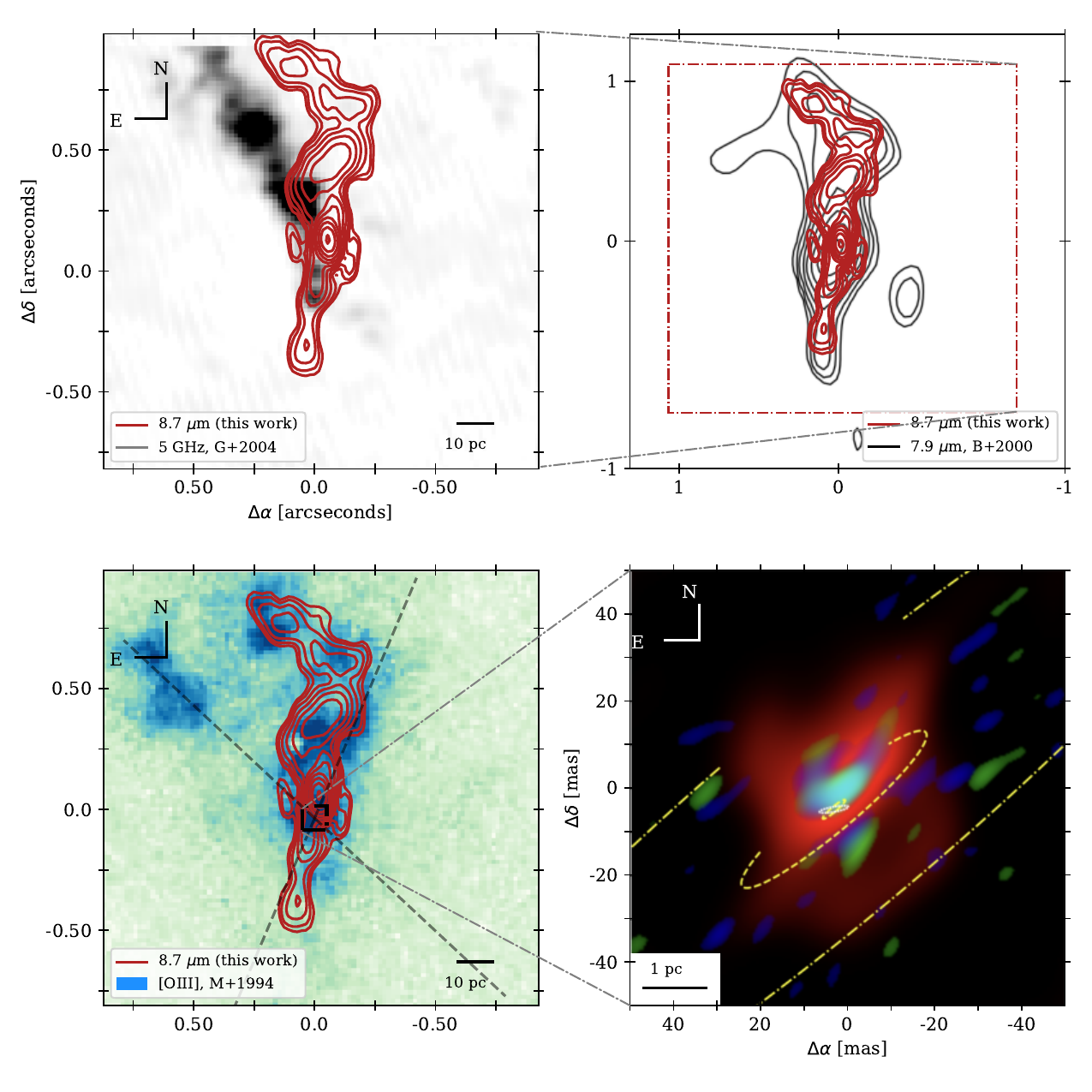}
\end{center}
\textbf{Supplementary Figure 3.} Comparison of CLEAN-deconvolved LBTI Fizeau image to observations at other wavelengths. Deconvolved LBTI Fizeau image of NGC 1068 at 8.7 $\mu$m compared to observations of the 5 GHz radio emission ({panel a}; [18]), the single-dish 7.9~\micron~emission ({panel b}; [29]), and the ionization cone traced by \oiii ~({panel c}; [28]). Panel d shows the MIR emission at sub-parsec scales as observed with VLTI/MATISSE (L-band in blue, M-band in green, and N-band in red; figure reproduced from [13]). The ``spur'' to the northeast of center found in the LBTI image is seen at the subparsec scale in the VLTI/MATISSE image. The yellow ellipses represent the orientation of the obscuring dust disk. The red contours are the same as in Fig. 2.

\clearpage
\begin{center}
\includegraphics[width=0.56\textwidth]{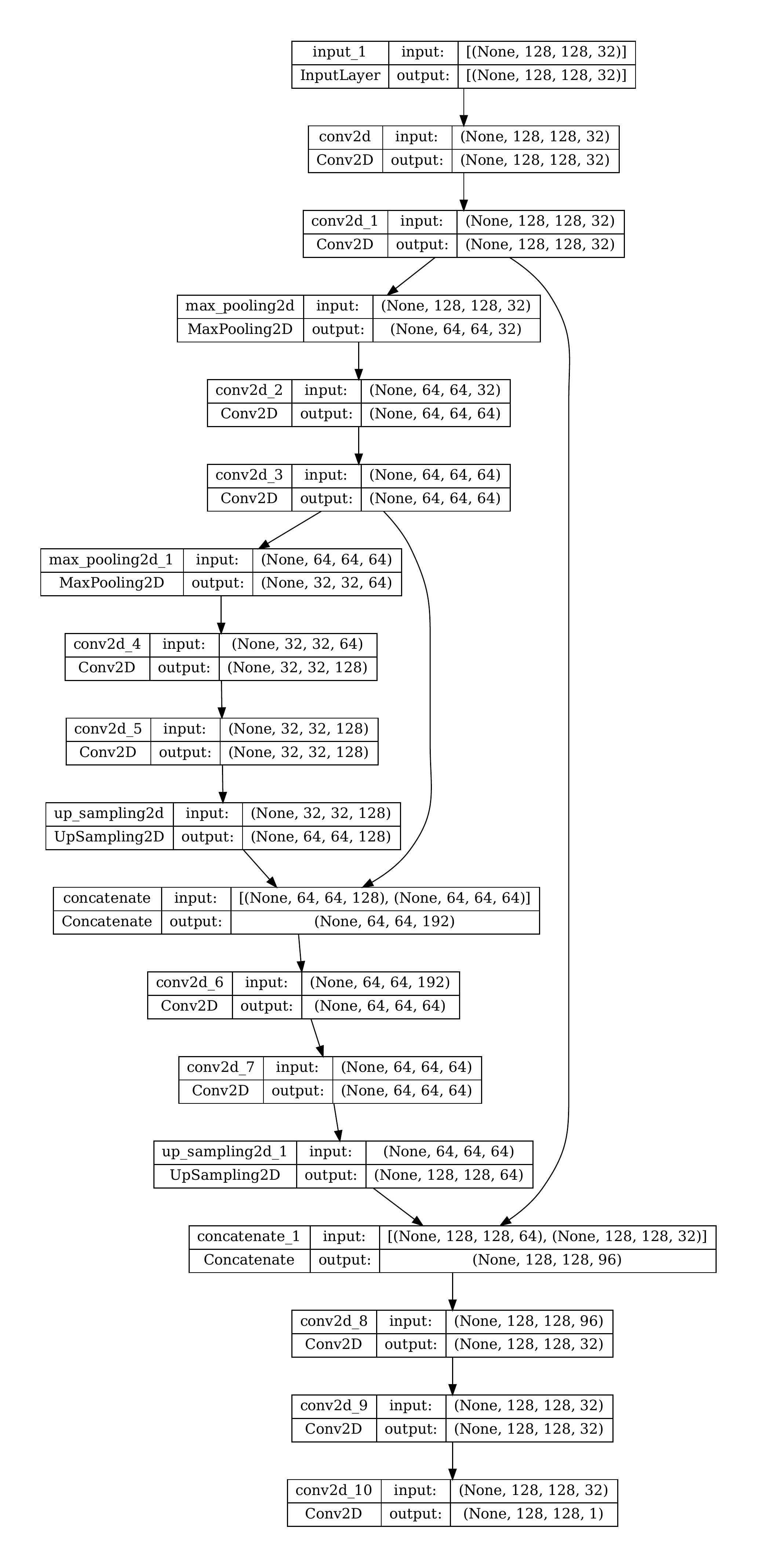}
\end{center}
\textbf{Supplementary Figure 4.} Schematic representation of the U-NET architecture used for deconvolution. Starting from an observation cube with dimensions ($n_{pix} \times n_{pix} \times n_{frames}$, the convolutional network concatenates layers both with and without MaxPool2D. The convolutions on different scales are then concatenated into a layer with dimensions ($n_{pix} \times n_{pix} \times 96$) before a final set of convolutions which produce the output image of dimension ($n_{pix} \times n_{pix}$).
\clearpage

\begin{center}
    \includegraphics[width=0.9\textwidth]{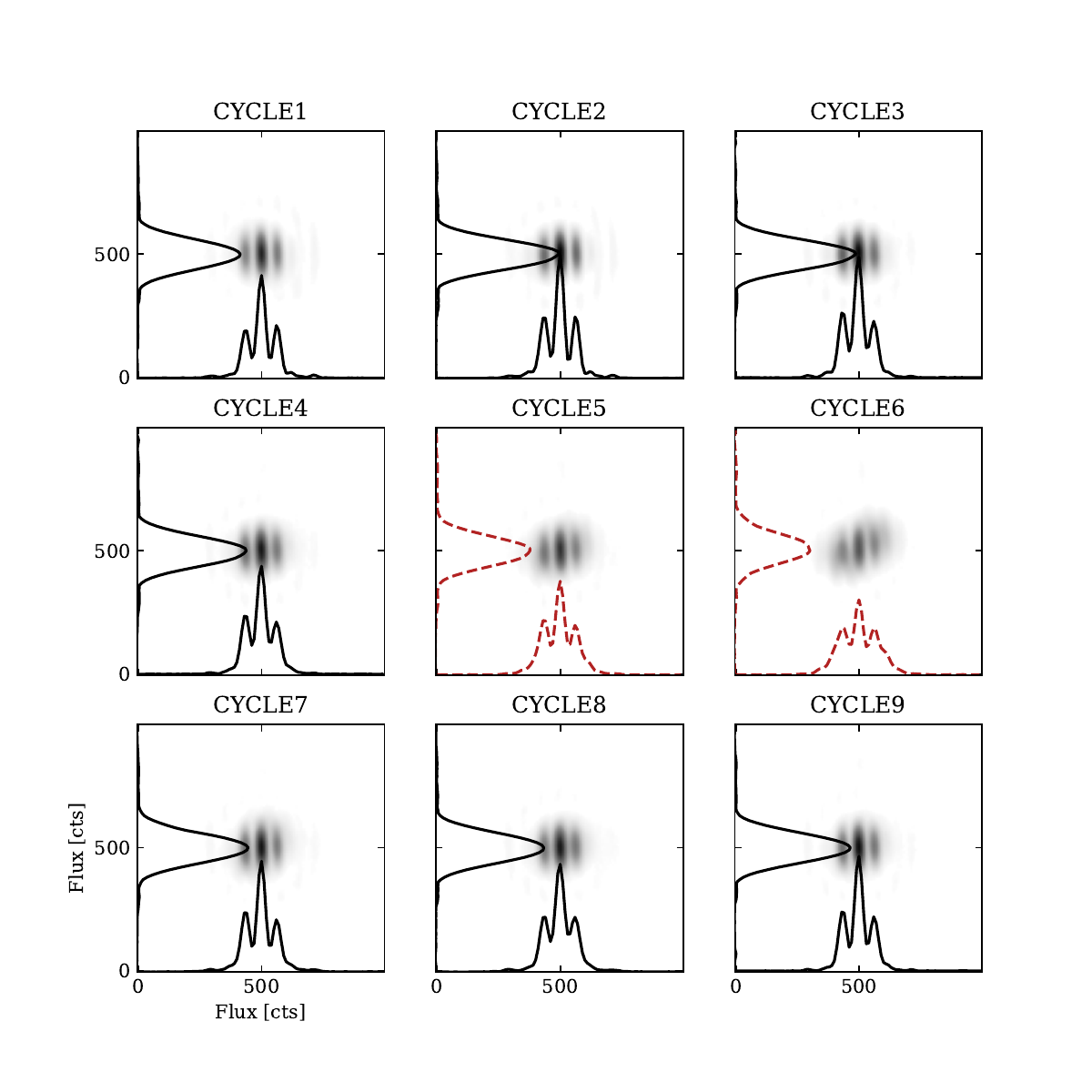}
\end{center}
\textbf{Supplementary Figure 5.} Individual PSF cycles of 1 Tau. Cycles 1 and 2 were obtained before the science target, and cycles 3-9 were obtained immediately after. Cycles 8 and 9 are presented in red because they are excluded from use in calibration due to cirrus clouds passing by during those exposures.
\clearpage


\begin{center}
\includegraphics[width=0.75\textwidth]{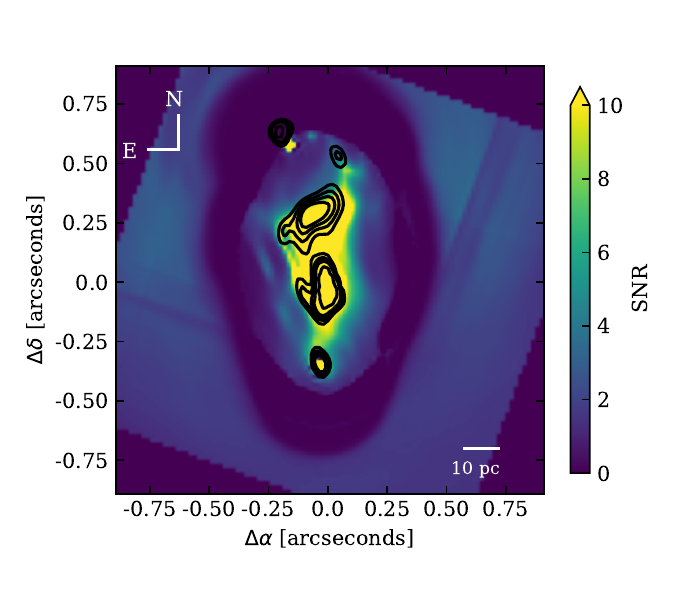}
\end{center}
\textbf{Supplementary Figure 6.} Bootstrapped SNR estimate from uncertainties in the PSF. The black contours are the same 8.7~$\mu$m image contours as in Figs. 2 and 3 ([2,3,4,5,10,20,50]$\times \sigma$). The high SNR region (in yellow) encloses all major features examined in this work.

\clearpage

\bibliographystyle{aasjournal}
\bibliography{main}

\end{document}